\documentclass[aps,prl,twocolumn,superscriptaddress,longbibliography]{revtex4-1}

\usepackage{amsmath}
\usepackage{graphicx}
\usepackage[colorlinks,linkcolor=blue,citecolor=blue,urlcolor=green]{hyperref}
\usepackage[dvipsnames]{xcolor}
\usepackage{stix}
\begin{document}

\newcommand{\dd}{\mathrm{d}}
\renewcommand{\vec}[1]{\mathbf{#1}}
\newcommand{\gvec}[1]{\boldsymbol{#1}}
\newcommand{\en}{\varepsilon}
\newcommand{\hc}{\hat{c}}
\newcommand{\hcd}{\hat{c}^\dagger}
\newcommand{\hd}{\hat{d}}
\newcommand{\hdd}{\hat{d}^\dagger}

\mathchardef\mhyphen="2A
\newcommand{\dx}[1]{\!\mathop{\mathrm{d}}\!#1\,}
\let\OldRe\Req
\renewcommand{\Re}{\text{Re}\,}
\let\OldIm\Im
\renewcommand{\Im}{\text{Im}\,}
\newcommand{\iu}{\mathrm{i}}

\newcommand{\intd}[1]{\int\!\!\dd #1\,}
\newcommand{\iintd}[2]{\int\!\!\dd #1\!\!\int\!\!\dd #2\,}

\newcommand{\toadd}[1]{\textcolor{cyan}{[#1]}}
\newcommand{\tocheck}[1]{\textcolor{red}{#1}}

\newcommand{\ph}{\mbox{$p$--$h$}}


\author{Michael Sch\"uler}
\affiliation{Department of Physics, University of Fribourg, 1700
  Fribourg, Switzerland}
\author{Jan Carl Budich}
\affiliation{Institute of Theoretical Physics, Technische Universit\"at Dresden, 01062 Dresden, Germany}
\author{Philipp Werner}
\affiliation{Department of Physics, University of Fribourg, 1700
  Fribourg, Switzerland}

\title{Quench Dynamics and Hall Response of Interacting Chern Insulators}

\begin{abstract}
  We study the coherent non-equilibrium dynamics of interacting
  two-dimensional systems 
  after a quench 
  from a
  trivial to a topological Chern insulator phase. While the many-body
  wavefunction is constrained to remain topologically trivial under
  local unitary evolution, we find that the Hall response of the
  system can dynamically approach a thermal value of the
  post-quench Hamiltonian, even though the efficiency of this
  thermalization process is shown to strongly depend on the microscopic form of
  the interactions. Quite remarkably, the effective temperature of the
  steady state Hall response can be arbitrarily tuned with the quench
  parameters. Our findings 
  suggest a new way of inducing and 
  observing low temperature topological phenomena in
  interacting ultracold atomic gases, where the considered quench
  scenario can be realized in current experimental set-ups.
\end{abstract}

\maketitle

Recent experimental progress in realizing topological insulators in a
non-equilibrium fashion in ultracold atomic
gases~\cite{Aidelsburger2013,Ketterle2013,jotzu_experimental_2014,
  wu_realization_2016-1,flaschner_experimental_2016,GoldmanReview2016}
and periodically driven solids~\cite{wang_observation_2013,mciver_light-induced_2018} 
raises fundamental questions regarding the quench
dynamics of topological phases. Rather than preparing the ground state
of the system of interest, the natural protocol in synthetic
material systems is to start from a trivial initial state, and quench
the (effective) Hamiltonian into a topological phase before observing the coherent
dynamics of the system. In this scenario topological
invariants of the many-body state, such as the Chern number, are conserved
(remain trivial), as the post-quench time-evolution
represents a local unitary
transformation~\cite{ChenLUT2010}.
Nevertheless, several observables provide signatures of a change in the
topological properties, including circular dichroism in
photoabsorption~\cite{tran_probing_2017-1,schuler_tracing_2017,tran_quantized_2018}, characteristic edge currents \cite{caio_quantum_2015}, and the
nonequilibrium Hall effect~\cite{dehghani_optical_2015,
  dehghani_out--equilibrium_2015,wang_universal_2016,hu_dynamical_2016,
schmitt_universal_2017-1,ulcakar_slow_2018,
peralta_gavensky_time-resolved_2018}. In
quenched noninteracting systems, the Hall response -- as the archetype
of a topological response property -- typically exhibits long-lasting
oscillations which may be reduced in certain cases by specifically
designed quench protocols~\cite{xu_scheme_2018}. In open
systems, where the aforementioned constraints on the temporal
invariance of topological properties are absent, the Hall response may
equilibrate due to extrinsic dephasing~\cite{hu_dynamical_2016} or
quantum
dissipation~\cite{wolff_dissipative_2016,schuler_tracing_2017}.

\begin{figure}[t]
  \includegraphics[width=\columnwidth]{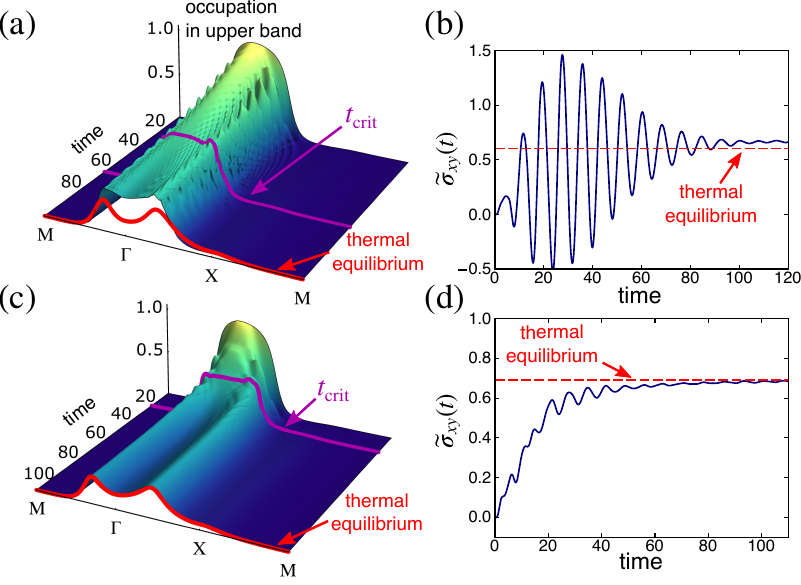}
  \caption{Thermalization of occupation and Hall response.
    Upper panels: dynamics of the occupation in the upper band
    of the post-quench Hamiltonian
    (a) and corresponding dynamical Hall response $\widetilde{\sigma}_{xy}(t)$ (b) in the interacting Chern
    insulator with local interactions only ($V=1$). Lower panels: occupation
    dynamics (c) and build-up of the Hall response (d) in the Chern insulator
    including nonlocal interactions ($V_0=1, V_1=0.25 V_0$). The results have been obtained
    within the GKBA, using a $N_k=220\times 220$ grid
    sampling of the BZ.
    \label{fig:toc}}
\end{figure}

The purpose of this work is to consider the quench dynamics of
{\textit{closed interacting}} topological 2D systems, where the time
evolution is still unitary at a global level, while two-body
scattering provides a source of {\textit{intrinsic dissipation}} which 
may enable thermalization processes. In this coherent scenario,
it is interesting to investigate which signatures of topology can
dynamically equilibrate despite the manifestly trivial character of
the time-evolved many-body state. Furthermore, thermalization in a
closed quantum system is a complex process, and, depending on the
allowed scattering processes, the system may be trapped in a
long-lived prethermal
state~\cite{berges_prethermalization_2004,moeckel_interaction_2008,
  eckstein_thermalization_2009,marcuzzi_prethermalization_2013,dalessio_quantum_2016}
in which local observables appear to be thermalized, whereas nonlocal
quantities are not. 
For 
a 1D system
it has recently been shown~\cite{kruckenhauser_dynamical_2017} that
the single-particle density matrix (SPDM) can thermalize towards an
equilibrium state of a topologically nontrivial post-quench
Hamiltonian. However, the dynamical equilibration of natural
observables for topological insulators, in particular 
{\textit{two-particle}} quantities such as the Hall conductivity in 2D
systems, has remained a largely unexplored question.

Below, we investigate the post-quench dynamics of the Hall response in
interacting 2D fermionic systems. With fully microscopic
numerical simulations based on the time-dependent nonequilibrium Green's functions (NEGF)
approach~\cite{stan_time_2009,balzer_nonequilibrium_2012,stefanucci_nonequilibrium_2013}, we demonstrate that a dynamical
equilibration to a {\textit{thermal}} value of the Hall response is
possible, if sufficiently many scattering channels are available,
while thermalization bottlenecks related to the topological
character of the considered quenches result in 
slow dynamics (see Fig.~\ref{fig:toc}). Specifically,
when considering a minimal model system for a Chern insulator, nearest
neighbor interactions are necessary to efficiently thermalize the Hall
response, even though onsite interactions already render the system
non-integrable.

\paragraph{Topological Hubbard model.--}
We study interacting topological insulator (TI) models on a 2D square lattice with unit lattice constant, defined by the Hamiltonian
$\hat{H} = \hat{H}_\mathrm{TI} + \hat{H}_\mathrm{int} $.
We consider two scenarios for the free Hamiltonian $\hat{H}_\mathrm{TI} $: (i) a spinless
Chern
insulator defined by
$\hat{H}_\mathrm{TI} = \sum_{\vec{k}}
\hat{\vec{c}}^\dagger_{\vec{k}} \vec{h}(\vec{k}) 
\hat{\vec{c}}_{\vec{k}}$ with
$\vec{h}(\vec{k})=(M-\cos(k_x)-\cos(k_y))\gvec{\sigma}_z + \lambda
\sin(k_x)\gvec{\sigma}_x + \lambda
\sin(k_y)\gvec{\sigma}_y$. Here, the $\hat{\gvec{c}}^{(\dagger)}_{\vec{k}} =
(\hat{c}^{(\dagger)}_{\vec{k}E}, \hat{c}^{(\dagger)}_{\vec{k}H})$ denote the fermionic
  annihilation (creation) operators with respect to the underlying two orbitals, labelled by E and H, respectively, in analogy to the original Bernevig-Hughes-Zhang (BHZ) \cite{bernevig_quantum_2006}
  model for HgTe, the $\gvec{\sigma}_i$ are the Pauli matrices
  in the orbital pseudospin space, and the lattice momentum ${\vec{k}}$ is defined in the first Brillouin zone (BZ).
  (ii) A spinful quantum spin Hall insulator 
  with time-reversal symmetry defined by \cite{bernevig_quantum_2006} $\hat{H}_\mathrm{TI} = \sum_{\vec{k},\sigma}
\hat{\vec{c}}^\dagger_{\vec{k},\sigma} \vec{h}_\sigma(\vec{k}) 
\hat{\vec{c}}_{\vec{k}\sigma}$ with $\vec{h}_\uparrow(\vec{k})
=\vec{h}(\vec{k})$,
$\vec{h}_\downarrow(\vec{k})=[\vec{h}(-\vec{k})]^*$. In both cases, the
system is a trivial band insulator for $M<-2$, while $-2<M<0$
corresponds to 
a (spin) Chern insulator with (spin) Chern number $C=1$. 
In what
follows, all energies (times) are measured in units of the hopping
(inverse hopping).

\paragraph{Post-quench dynamics with interactions.--}

To gain insights into the dynamical manifestation of topological
properties, the system is prepared in a low-temperature equilibrium
state in the topologically trivial phase. At times $t=0_+$, the mass
parameter $M$ is suddenly switched (continuous ramps of $M$ will be
considered further below) to the topological regime and kept constant
for $t>0$. To disentangle scattering processes in the post-quench
dynamics from the initial state, $\hat{H}_\mathrm{int}$ is also
switched on suddenly at $t=0_+$. 
This protocol can be realized 
experimentally in ultracold atomic gases by tuning a magnetic field in
the vicinity of a Fano-Feshbach
resonance~\cite{QuenchWeakInteraction2013,QuenchUnitary2014}.  We will
consider weak to intermediate inter-particle interactions in this
work, so that the topological character of the system is determined by
$\hat{H}_\mathrm{TI}$.

The time-dependent NEGF
approach~\cite{stan_time_2009,balzer_nonequilibrium_2012,stefanucci_nonequilibrium_2013,
  aoki_nonequilibrium_2014} will be used to describe the
\emph{correlated} dynamics. This method is based on solving the
Kadanoff-Baym equations (KBEs) for the single-particle Green's
function (SPGF), from which the
SPDM $\gvec{\rho}_\sigma(\vec{k};t)$ and thus all
single-particle observables can be computed. The KBEs yield, in
principle, the exact SPGF, provided the self-energy kernel is
known. In practice however, additional approximations are inevitable.
In this work, we employ the second-Born approximation (2BA) to the
self-energy, which corresponds to a second-order expansion in the
two-body interaction. The 2BA has been shown to provide an excellent
description of the electronic structure and dynamics for relatively weak
interactions~\cite{balzer_stopping_2016,schuler_spectral_2018}.

In order to extrapolate to the thermodynamic limit, the number of
points $\vec{k}$ sampling the BZ has to be chosen
sufficiently large, which -- due to the substantial numerical effort
of solving the full KBEs -- poses a computational
challenge. 
Invoking the generalized Kadanoff-Baym ansatz
(GKBA)~\cite{lipavsky_generalized_1986} is an additional approximation
which reduces  the numerical effort significantly~\cite{Note1}.
The GKBA has been shown to yield excellent results in the
weak-coupling regime for single-particle observables~
\cite{latini_charge_2014,perfetto_first-principles_2015,balzer_stopping_2016,
perfetto_ultrafast_2018,bostrom_charge_2018}.
How well nonequilibrium response properties are captured is less
understood and will be addressed in the context of the Hall response
below.
%
Both the full KBE and the GKBA treatment conserve the total
energy. Therefore, comparing the energy of the system after the
quench to the thermal equilibrium energies of post-quench interacting
system allows to determine the effective temperature
$T_\mathrm{eff}$ and corresponding thermalized observables.

\paragraph{Chern insulator with local interactions.--} 

As the first paradigmatic example we consider the case where 
$\hat{H}_\mathrm{TI}$ defines a spinless Chern insulator for
$M<-2$. Restricting to local interactions, we consider the interaction term 
\begin{align}
  \hat{H}_\mathrm{int} =
  \frac{V}{2}\sum_{i}\sum_{\alpha\ne\alpha^\prime}
  \hat{n}_{i,\alpha}\hat{n}_{i,\alpha^\prime} \ ,
\end{align}
where $i$ runs over all lattice sites, while
$\alpha,\alpha^\prime\in\{E,H\}$. In the following, we fix
$\lambda=0.4$ and consider the quench of the gap parameter
$M_\mathrm{pre}\rightarrow M_\mathrm{post}$ with 
$M_\mathrm{pre}=-3.5$ and $M_\mathrm{post}=-1$. 

Without band hybridization ($\lambda=0$), the $E$ and $H$ bands
possess a $U(1)$ symmetry, which results in individually conserved
particle numbers $n_E$ and $n_H$. As a result, inter-orbital (and thus
inter-band) thermalization will be completely suppressed. In the case
$\lambda>0$ and $C=1$, inter-orbital scattering in the
lower and upper band becomes possible, albeit only active close to the
avoided crossings. Furthermore, the non-vanishing Chern number
implies that even within the same Bloch band, there are
states with opposite orbital character 
which are not connected by the intra-orbital interaction. 
Therefore, thermalization can only proceed via higher-order
scattering processes.

The expectation of slow thermalization is confirmed by inspecting the time evolution of
the occupation in the upper band $f_+(\vec{k};t) = 
\gvec{\phi}^\dagger_{\vec{k},+} \gvec{\rho}(\vec{k};t) \gvec{\phi}_{\vec{k},+}$ with respect to the
post-quench free Hamiltonian $\vec{h}(\vec{k}) \gvec{\phi}_{\vec{k},\pm}
= \varepsilon_\pm(\vec{k}) \gvec{\phi}_{\vec{k},\pm}$, presented in
Fig.~\ref{fig:toc}(a) along the path M--$\Gamma$--X--M in the BZ for
$V=1.0$.  Initially, the SPDM is prepared as the equilibrium
state of the band insulator with dominant $E$ orbital character;
quenching $M$ leads to a band inversion with preserved occupation of
the $E$ orbital, which after the quench has large weight in the upper band near the center
of the BZ. The subsequent relaxation due to
particle-particle scattering reduces the number of excited
carriers. Computing the injected energy and comparing to thermal
equilibrium yields the effective temperature $T_\mathrm{eff}$. The
corresponding equilibrium occupation
$f^\mathrm{eq}_+(\vec{k};T_\mathrm{eff})$ is represented by the red
line in Fig.~\ref{fig:toc}(a). The deviation of $f_+(\vec{k};t)$ from
$f^\mathrm{eq}_+(\vec{k};T_\mathrm{eff})$ illustrates 
the very slow approach to thermal equilibrium after the quench. 
Furthermore, the oscillations of the occupation indicate a
coherent superposition of the lower and upper band, which is only slowly
damped; hence, dephasing exists but is ineffective.

\begin{figure}[t]
  \includegraphics[width=\columnwidth]{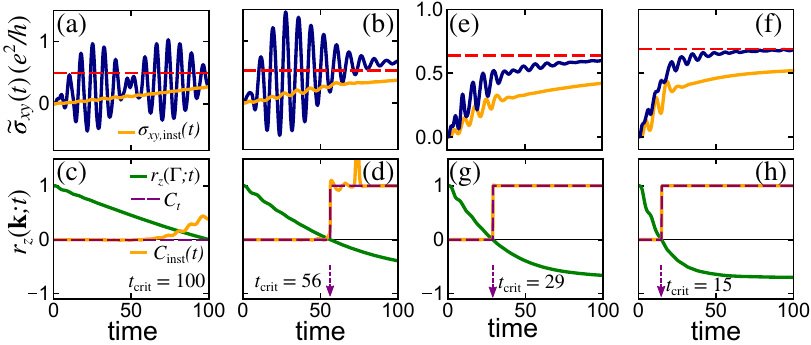}
  \caption{Dynamics of the Hall response and orbital pseudospin for a
    Chern insulator with local interactions only ((a)--(d)) and with
    nonlocal interactions ((e)--(h)). The interaction strength
    is $V=0.65$ ((a) and (c)) and $V=1.0$ ((b) and (d))
    in the local case, while for the model with nonlocal interactions it is
    $V_0=0.65$ ((e) and (g)) and $V_0=1.0$ ((f) and (h)). We used the
    GKBA and show results for $N_k=220\times 220$ grid
    points in the BZ. \label{fig:chern}}
\end{figure}

To probe the dynamical Hall response $\widetilde{\sigma}_{xy}(t)$, we 
apply a weak electric field $E_y(t)=F_0(1-e^{-t/\tau})$ in the 
$y$-direction ($F_0=10^{-3}$ and $\tau=5$). 
Measuring the induced current in the $x$-direction $J_x(t)$ yields the Hall
response $\widetilde{\sigma}_{xy}(t) =
J_x(t)/F_0$. Figure~\ref{fig:chern} shows $\widetilde{\sigma}_{xy}(t)$
for $V=0.65$ (a) and $V=1.0$ (b) [also shown in
Fig.~\ref{fig:toc}(b)]. For $V=0.65$, strong coherent oscillations around
a nonthermal value of the Hall conductance dominate and the relaxation
to a steady state $\widetilde{\sigma}_{xy}(t) \rightarrow
\sigma_{xy}$ ($t\rightarrow\infty$) cannot be
observed on the accessible timescales. Increasing the interaction strength to $V=1.0$, a steady
state begins to form at times $t\approx 100$; however, its Hall conductance
deviates from the thermal equilibrium value, which is consistent with 
the nonthermal distribution (Fig.~\ref{fig:toc}(a)). 

Further insight into the topological properties of the SPDM can be gained by
extracting the pseudospin vector $\vec{r}(\vec{k;t})$
via the relation
$\gvec{\rho}(\vec{k};t)=[\gvec{I}-\vec{r}(\vec{k;t})\cdot
\gvec{\sigma}]/2$. As long as $\vec{r}(\vec{k;t}) \ne 0$, the SPDM can formally be expressed as
$\gvec{\rho}(\vec{k};t) =\exp(-\vec{h}_\mathrm{aux}(\vec{k};t))$ with
a \emph{gapped} auxiliary Hamiltonian
$\vec{h}_\mathrm{aux}(\vec{k};t)$. Closing of the gap of
$\vec{h}_\mathrm{aux}(\vec{k};t)$, known from open systems as a purity
gap closing
\cite{Diehl2011,BardynPRX2018},
marks a dynamical topological transition at the
critical time $t_\mathrm{crit}$ corresponding to 
$\vec{r}(\Gamma;t_\mathrm{crit})=0$~\cite{Note2}.
The occurrence of the band inversion indicated by
$r_z(\Gamma;t)$ passing through zero also
allows to define a simplified topological index $C_t$ via
$(-1)^{C_t} = \mathrm{sign}(r_z(\Gamma;t)r_z(X;t))$~\cite{Note3}.
This index is plotted together with $r_z(\Gamma;t)$ in
Fig.~\ref{fig:chern}(c)--(d). One finds a purity gap closing at
$t_\mathrm{crit}\approx 100$ for $V=0.65$ and
$t_\mathrm{crit}\approx 56$ for $V=1.0$. Fig.~\ref{fig:toc}(a) shows
the distribution $f_+(\vec{k};t_\mathrm{crit})$, which changes its
curvature at $\vec{k}=\Gamma$ at $t=t_\mathrm{crit}$, indicating a band
inversion.
Further
analysis~\cite{supplement} shows a decrease of 
$t_ \mathrm{crit}$ proportional to $V^{-3/2}$.

The pseudospin structure furthermore allows to define the
instantaneous Berry curvature 
\begin{align}
  \label{eq:berry}
  \Omega(\vec{k};t) = -\frac12 \hat{\vec{r}}(\vec{k};t)\cdot
  \left(\frac{\partial\hat{\vec{r}}}{\partial k_x} \times
  \frac{\partial\hat{\vec{r}}}{\partial k_y}\right) \ ,
\end{align}
where
$\hat{\vec{r}}(\vec{k};t) =
\vec{r}(\vec{k};t)/|\vec{r}(\vec{k};t)|$. The Berry
curvature~\eqref{eq:berry} defines the instantaneous Chern number of the SPDM
$C_\mathrm{inst}(t)=(1/2\pi)\int_{\mathrm{BZ}}\dd\vec{k}\,
\Omega(\vec{k};t)$ and the instantaneous Hall conductance
$\sigma_{xy,\mathrm{inst}}(t)=\int_{\mathrm{BZ}}\dd\vec{k}\,
|\vec{r}(\vec{k};t)|\Omega(\vec{k};t)$. Importantly,
$C_\mathrm{inst}(t)$ would be pinned to zero in a noninteracting
system. In contrast, Fig.~\ref{fig:chern}(c)--(d) shows a nonzero
instantaneous Chern number, which for $V=1.0$ becomes almost identical
to $C_t$~\cite{Note4}.
The instantaneous
Hall conductance exhibits a fast increase for $t<t_\mathrm{crit}$ and a saturation after
$t>t_\mathrm{crit}$.
However, $\sigma_{xy,\mathrm{inst}}(t)$ does not coincide with the
thermal Hall conductance of the interacting system.

\paragraph{Chern insulator with nonlocal interactions.--} 
The thermalization process changes substantially if nonlocal interactions are
included. For the purpose of this study, we consider
\begin{align}
  \hat{H}_\mathrm{int} =
  \frac{1}{2}\sum_{i,j}\sum_{\alpha,\alpha^\prime}
  V^{\alpha\alpha^\prime}_{i,j}\hat{n}_{i,\alpha}\hat{n}_{i,\alpha^\prime} \ .
\end{align}
Here, the interactions are $V^{\alpha\alpha^\prime}_{i,i} =
V_0(1-\delta_{\alpha\alpha^\prime})$ and
$V^{\alpha\alpha^\prime}_{i,j} =V_1$ if $i$ and $j$ are nearest
neighbours. The nonlocal repulsion is fixed to $V_1=0.25 V_0$.

Figure~\ref{fig:toc}(c) depicts the occupation of the upper band (with
respect to the post-quench free Hamiltonian) $f_+(\vec{k};t)$ for
$V_0=1.0$. In contrast to the model with local interactions only, the
nonlocal part of the interaction includes inter-orbital and
intra-orbital scattering, which results in a rapid thermalization of
$f_+(\vec{k};t)$ to the equilibrium
$f^\mathrm{eq}_+(\vec{k})$. Furthermore, coherent oscillations are
suppressed, indicating pronounced dephasing. The nonequilibrium Hall
response is shown in Fig.~\ref{fig:chern}(e)--(f) for $V_0=0.65$ and
$V_0=1.0$, respectively. In this set-up, $\widetilde{\sigma}_{xy}(t)$
approaches the thermal equilibrium value within the numerically
accessibe time window. Furthermore, it shows a qualitatively very similar behavior as
the instantaneous conductance $\sigma_{xy,\mathrm{inst}}(t)$,
indicating strong dephasing. There are two different regimes: a
rapid increase with superimposed oscillations for $t<t_\mathrm{crit}$ and a
smooth saturation for $t>t_\mathrm{crit}$. This behavior of
$\sigma_{xy,\mathrm{inst}}(t)$ is also reflected in 
$\widetilde{\sigma}_{xy}(t)$. The time scale of the purity gap closing
(Fig.~\ref{fig:chern}(g)--(f)) is significantly shorter as compared to
the case with local interactions only: $t_\mathrm{crit}\approx 29$ for
$V_0=0.65$ and $t_\mathrm{crit}\approx 15$ for $V_0=1.0$. The sign of
the curvature of the distribution $f_+(\Gamma;t)$ changes at
$t=t_\mathrm{crit}$ (Fig.~\ref{fig:toc}(c)). 
Again, $t_\mathrm{crit}$ scales as $V^{-3/2}$.

\paragraph{Quantum spin Hall insulator.--}
Including the spin degree of freedom while requiring time-reversal
symmetry gives rise to a $\mathbb{Z}_2$ quantum spin Hall insulator. 
This is the typical scenario in materials where the
spin-orbit interaction is the mechanism behind the topological gap
opening. In this case, the on-site Hubbard repulsion (which
is excluded by the Pauli principle in the spinless case) becomes the simplest
possible interaction term~\cite{budich_fluctuation-driven_2013}. For the
sake of consistency with the previous discussion, we also include a local
inter-orbital coupling and define the interaction term as
\begin{align}
  \hat{H}_\mathrm{int} = U \sum_{i,\alpha}
  \hat{n}_{i,\alpha,\uparrow}\hat{n}_{i,\alpha,\downarrow} +
  \frac{V}{2}\sum_{i,\sigma}\sum_{\alpha\ne\alpha}\hat{n}_{i,\alpha,\sigma}
\hat{n}_{i,\alpha^\prime,\sigma}.
\end{align}
We employ the same quench protocol as
above and fix $M_\mathrm{pre}=-3.5$ and $M_\mathrm{post}=-1.5$.

\begin{figure}[t]
  \includegraphics[width=\columnwidth]{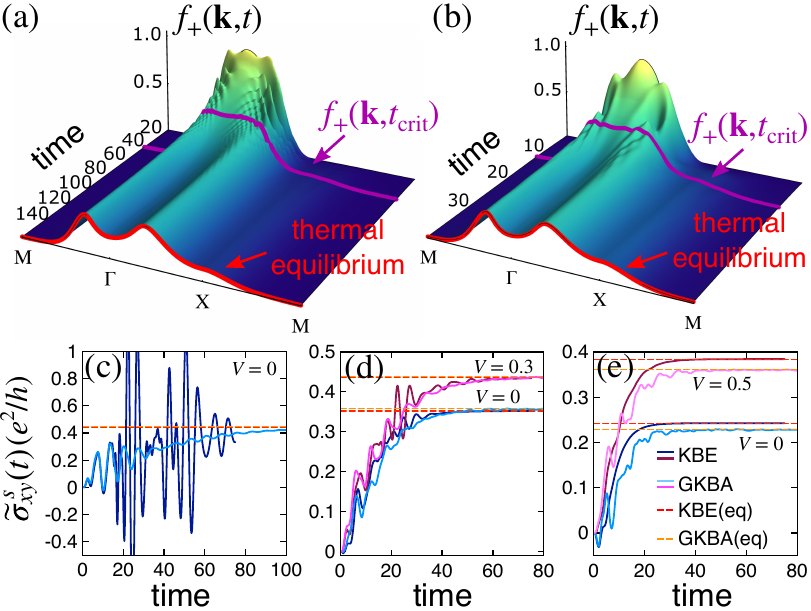}
  \caption{Dynamics of the occupation $f_+(\vec{k};t)$ in the upper
    band of a quantum spin Hall insulator
    for (a) $U=1.0, V=0$, and (b) $U=1.5, V=0$. The results have
    bween obtained employing the GKBA. The lower panels
    depict the
    nonequilibrium spin Hall conductance for (c) $U=1.0$, (d)
    $U=1.5$, and (e) $U=2.0$. \label{fig:z2hall}
    }
\end{figure}

Inspecting the occupation in the upper band for weak ($U=1.0$, Fig.~\ref{fig:z2hall}(a))
 and slightly increased ($U=1.5$, Fig.~\ref{fig:z2hall}(b))
Hubbard repulsion, one finds
rapid dephasing and thermalization. Hence, the on-site
intra-orbital interaction is sufficient to fully thermalized the
system on short time scales (determined by the interaction
strength). Adding the inter-orbital coupling $V$ does not change this
behavior, albeit the thermalization becomes slightly slower. This is
a mean-field effect: the inter-orbital interaction increases the
effective post-quench gap parameter $M^\mathrm{eff}_\mathrm{post} =
M_\mathrm{post} -U n_{E} - 2V n_H + V \rho_{EH}$, where $\rho_{EH}$
denotes the off-diagonal element of the local density matrix. This
Fock term thus leads to an effectively reduced quench size
$M^\mathrm{eff}_\text{post}-M_{\mathrm{pre}}$, which injects less energy
into the system, thus giving rise to slower thermalization. A
systematic analysis for different $M_{\mathrm{pre}}$ and
$M_{\mathrm{post}}$ ~\cite{supplement} confirms this picture.

While the total Hall conductance vanishes in the spin Chern insulator, 
the spin Hall conductance
$\widetilde{\sigma}^s_{xy}(t) = [\widetilde{\sigma}^{\uparrow}_{xy}(t)
-\widetilde{\sigma}^{\downarrow}_{xy}(t)]/2$ becomes quantized to one
at zero temperature, due to the presence of a $U(1)$ spin rotation symmetry. The spin Hall conductance is presented in
Fig.~\ref{fig:z2hall}(c)--(e). Within the GKBA on a
$N_k=200\times 200$ grid, which corresponds to the converged
thermodynamic limit in all cases, $\widetilde{\sigma}^s_{xy}(t)$
rises rapidly and approaches the thermal equilibrium value at the
corresponding effective temperature $T_\mathrm{eff}$. As for the Chern
insulator, the characteristic time scale for the build-up of the spin Hall effect is
the critical time $t_\mathrm{crit}$ of the purity gap
closing (also indicated in Fig.~\ref{fig:z2hall}(a)--(b)). Increasing the 
strength of the Hubbard repulsion (while keeping $V=0$) leads to
significantly enhanced dephasing, while the steady-state spin Hall
conductance $\sigma^s_{xy}$ is reduced. This can again be attributed to the increase
of injected energy due to a stronger effective quench. Including the 
inter-orbital interaction $V$ counter-acts this effect and thus
results in a larger 
$\sigma^s_{xy}$. 
Hence, tuning the inter-orbital coupling $V$
provides a way of effectively cooling down the system and thus
increasing $\sigma^s_{xy}$.

It is interesting to compare the GKBA to the full solution of the
KBEs (darker lines in Fig.~\ref{fig:z2hall}(c)--(e)). Due to the
numerical effort, the KBE simulations are limited to a $32\times 32$
cluster here. For $U=1.0$, the GKBA and KBE results agree well up to
$t\approx 15$; for later times, finite-size effects dominate the KBE
dynamics. Nevertheless, the steady-state $\sigma^s_{xy}$ agrees
well. For the slightly larger interaction $U=1.5$
(Fig.~\ref{fig:z2hall}(d)), where dephasing is significantly enhanced and
finite-size effects thus suppressed, the GKBA and KBE dynamics
agree well, as does the steady state. For stronger interaction, however, 
deviations become apparent (Fig.~\ref{fig:z2hall}(e)).  The KBE
dynamics (which has already converged to the thermodynamic limit)
shows much more pronounced dephasing and smoothly approaches the
thermal steady state. While the GKBA is also thermodynamically
consistent, it seems to underestimate the KBE Hall
conductance. Analyzing the Hall conductance in thermal
equilibrium~\cite{supplement}, we find that the GKBA fails to
reproduce the integer Hall effect for $T\rightarrow 0$~\cite{Note5}.
Apart from this subtlety, for moderate interaction strength
$U\lesssim 1.5$, the GKBA and KBE treatments agree very well. Since the
exact solution is typically in
between~\cite{schlunzen_nonequilibrium_2017}, this comparison establishes
the GKBA as an excellent method in the weak-interaction regime, and demonstrates
the predictive power of our approach.

\begin{figure}[t]
  \includegraphics[width=\columnwidth]{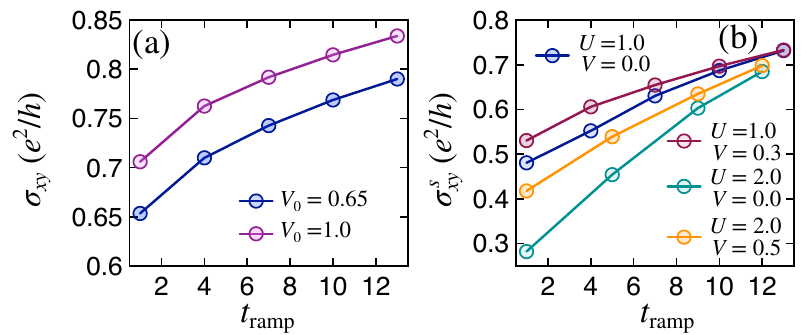}
  \caption{Steady-state (spin) Hall conductance of (a) the spinless
    Chern insulator with nonlocal interactions ($V_1=0.25 V_0$), and
    (b) the spinfull $\mathbb{Z}_2$ insulator as a function of the
    ramp time. The GKBA has been used in (a) and (b) for all results,
    except for $U=2.0$, where the KBE result is
    shown. \label{fig:hall_ramp}}
\end{figure}

\paragraph{Realizing a low-temperature topological state.--} 
The largest steady-state values $\sigma_{xy}$ ($\sigma^s_{xy}$) for the
Chern (spin Chern) insulator are around $\sigma_{xy}\simeq 0.68$ and
$\sigma^s_{xy}\simeq 0.41$, respectively, corresponding to a
high-temperature state after thermalization. The effective temperature
can, however, be lowered significantly by slowly \emph{ramping} $M$. 
We demonstrate this behavior by modifying our
protocol to: (i) preparing the band-insulating system in equilibrium
including $\hat{H}_\mathrm{int}$, and (ii) modifying the gap
according to $M(t) = M_\mathrm{pre} + (M_\mathrm{post}  -
M_\mathrm{pre} ) f(t/t_\mathrm{ramp})$~\cite{Note6}.
$t_\mathrm{ramp}$ defines the ramp duration ($M(t_\mathrm{ramp})=M_\mathrm{post}$).
Figure~\ref{fig:hall_ramp}(a) ((b)) shows the steady-state (spin) Hall
conductance for various values of the interaction strength. As
Fig.~\ref{fig:hall_ramp} demonstrates, the thermal Hall effect can be
significantly enhanced for all interactions by increasing
$t_\mathrm{ramp}$. Slower ramps lead to an adiabatic time evolution for more and more
points in the BZ except for the region close to the gap closing at the
$\Gamma$ point. Therefore, the ability of the system to thermalize results
in an arbitarily low $T_\mathrm{eff}$ and a $\sigma_{xy}$ or
$\sigma^s_{xy}$  approaching one. This, however, would not be the case
for the nonthermal Chern insulator with local interactions only.

\paragraph{Conclusions.--}
We have systematically investigated the post-quench
dynamics of closed interacting two-dimensional topological
insulators. We showed that the system can dynamically approach thermal equilibrium
with respect to the topological properties of the SPDM \emph{and} the Hall response, even though the
topological invariant of the many-body state stays pinned to the trivial
value. Thus, our results demonstrate that the eigenstate
thermalization hypothesis also applies to topologically constrained
unitary time evolution. The microscopic scattering mechanisms play a crucial role: 
while the spin-less Chern insulator with local
interactions only is non-integrable, it thermalizes very slowly 
due to topological restrictions. In contrast, including
intra-orbital coupling by nonlocal interactions or extending to a
spinfull $\mathbb{Z}_2$ topological insulator
accelerates the 
thermalization on the one-
and two-particle level. Therefore, switching from the topologically trivial
to the nontrival regime by slow ramps allows to realize
a low-temperature state with almost integer Hall
conductance, providing a new way of dynamically inducing and
observing topological phenomena.

\begin{acknowledgements}

This work was supported by the Swiss National Science Foundation
through NCCR MARVEL and the European Research Council through ERC
Consolidator Grant 724103. The calculations have been performed on the
Beo05 cluster at the University of Fribourg, and the Piz Daint cluster
at the Swiss National Supercomputing Centre (CSCS).

\end{acknowledgements}

\appendix

\section{Appendix: Theoretical methods}

\subsection{Time-dependent nonequilibrium Green's function approach}

The time-dependent nonequilibrium Green's function (TDNEGF) method is
based on solving the Kadanoff-Baym equations (KBEs) for the
single-particle Green's function (SPGF):
\begin{align}
  \left(\iu \partial_t - \vec{h}(\vec{k}; t) \right) \vec{G}(\vec{k};
  t,t^\prime) = \delta_\mathcal{C}(t,t^\prime) +
  \int_{\mathcal{C}}\!\dd \bar{t}\, \gvec{\Sigma}(\vec{k};t,\bar{t})
  \vec{G}(\vec{k};\bar{t},t^\prime) \ .
\end{align}
The time arguments of the SPGF $\vec{G}(\vec{k}; t,t^\prime)$ lie on
the L-shaped Kadanoff-Baym contour $\mathcal{C}$. In practice, 
we solve this equation 
by introducing 
a set of 
two-time correlators~\cite{stefanucci_nonequilibrium_2013}. 
The resulting KBEs 
are
\begin{subequations}
  \label{eq:kbes}
  \begin{equation}
    \iu \partial_t \vec{G}^>(\vec{k}; t,t^\prime) = \vec{h}(\vec{k};t)
    \vec{G}^>(\vec{k}; t,t^\prime) 
    + \left[\gvec{\Sigma}(\vec{k})\ast
      \vec{G}(\vec{k})\right]^>(t,t^\prime) \ , 
  \end{equation}
  \begin{equation}
    -\iu \partial_t \vec{G}^<(\vec{k}; t^\prime,t) = 
    \vec{G}^>(\vec{k}; t^\prime, t) \vec{h}(\vec{k};t)
    + \left[
      \vec{G}(\vec{k})\ast \gvec{\Sigma}(\vec{k})\right]^<(t^\prime,t) \ , 
  \end{equation}
  \begin{equation}
    \label{eq:kbe_tv}
    \iu \partial_t \vec{G}^\rceil(\vec{k}; t,\tau) = 
    \vec{h}(\vec{k};t)\vec{G}^\rceil (\vec{k}; t, \tau) 
    + \left[
      \gvec{\Sigma}(\vec{k})\ast\vec{G}(\vec{k})\right]^\rceil(t,\tau) \ .
  \end{equation}
\end{subequations}
Here, the standard Langreth rules~\cite{stefanucci_nonequilibrium_2013} define the convolution
\begin{align*}
    \left[
      \vec{A}(\vec{k})\ast
      \vec{B}(\vec{k})\right]^\gtrless(t,t^\prime)  &= \int^t_0\!\dd
  \bar{t}\, \vec{A}^\mathrm{R}(\vec{k}; t,\bar{t})
  \vec{B}^\gtrless(\vec{k};\bar{t},t^\prime) \\&\quad + \int^{t^\prime}_0\!\dd
  \bar{t}\, \vec{A}^\gtrless(\vec{k}; t,\bar{t})
  \vec{B}^\mathrm{A}(\vec{k};\bar{t},t^\prime) \\ &\quad -\iu \int^\beta_0\!\dd
                                                 \tau\,
                                                 \vec{A}^\rceil(\vec{k};
                                                 t,\tau) \vec{B}^\lceil(\vec{k};
                                                 \tau,t^\prime)  
\end{align*}
and 
\begin{align*}
 \left[
      \vec{A}(\vec{k})\ast
      \vec{B}(\vec{k})\right]^\rceil(t,\tau)  &= \int^t_0\!\dd
  \bar{t}\, \vec{A}^\mathrm{R}(\vec{k}; t,\bar{t})
  \vec{B}^\rceil(\vec{k};\bar{t},\tau) \\ &\quad 
                                            +\int^\beta_0\!\dd
                                            \tau^\prime\,
                                            \vec{A}^\rceil(\vec{k};
                                            t,\tau^\prime)
                                            \vec{B}^\mathrm{M}(\vec{k};\tau^\prime-\tau)
                                            \ . 
\end{align*}
For a given choice of the many-body self-energy
$\gvec{\Sigma}(\vec{k})$, one first obtains for the Matsubara SPGF
$\vec{G}^\mathrm{M}(\vec{k};\tau)$ which captures initial
correlations. With the initial conditions thus determined, the
KBEs~\eqref{eq:kbes} govern the real-time evolution. 

For the quench setup employed the the main text, the
KBEs~\eqref{eq:kbes} simplify due to the lack of initial correlations,
leading to $\gvec{\Sigma}^\rceil(\vec{k};t,\tau) =0$. In this
scenario, the initial conditions (keeping track of orbital and spin indices) are determined by
\begin{align}
  \label{eq:initialcond}
  G^<_{\alpha\alpha^\prime \sigma}(\vec{k}; 0,0) &= \iu
  \widetilde{\rho}_{\alpha \alpha^\prime\sigma}(\vec{k}) \ , \nonumber
  \\ G^>_{\alpha\alpha^\prime \sigma}(\vec{k}; 0,0) &= -\iu
  (\delta_{\alpha \alpha^\prime} - \widetilde{\rho}_{\alpha
  \alpha^\prime\sigma}(\vec{k}) ) \ ,
\end{align}
where $\widetilde{\rho}_{\alpha \alpha^\prime\sigma}(\vec{k})$ is the density matrix
corresponding to the (uncorrelated) pre-quench equilibrium state.

In either setup, the KBEs~\eqref{eq:kbes} are solved with an in-house
massively-parallel computer code (used also in
Ref.~\cite{schuler_nonthermal_2018}) based on a fifth-order
predictor-corrector scheme. An equidistant time step of $\Delta t =
0.05$ was used, ensuring the convergence of all observables.

\subsection{Generalized Kadanoff-Baym ansatz}

The generalized Kadanoff-Baym ansatz
(GKBA)~\cite{lipavsky_generalized_1986} reduces the
KBEs~\eqref{eq:kbes} to an equation of motion for the density matrix
\begin{align}
  \frac{\dd}{\dd t}\gvec{\rho}(\vec{k};t) = -\iu
  \left[\vec{h}^{\mathrm{MF}}(t), \gvec{\rho}(\vec{k};t) \right]
  - \vec{I}(\vec{k},t) \ ,
\end{align}
where the collision term $\vec{I}(\vec{k},t)$ is defined by
\begin{align}
  \label{eq:collint}
  \vec{I}(\vec{k},t) =
  [\gvec{\Sigma}(\vec{k})\ast\vec{G}(\vec{k})]^<(t,t) + \mathrm{h.\,
  c.} \ .
\end{align}
The time off-diagonal SPGF required for computing the collision
integral~\eqref{eq:collint} are reconstructed by the GKBA
\begin{align}
  \label{eq:gkba1}
  -\iu \vec{G}^\gtrless(\vec{k};t,t^\prime) &=
  \vec{G}^\mathrm{R}(\vec{k};t,t^\prime)
  \vec{G}^\gtrless(\vec{k};t,t^\prime) \nonumber  \\ &\quad  -
  \vec{G}^\gtrless(\vec{k};t,t^\prime)
  \vec{G}^\mathrm{A}(\vec{k};t,t^\prime) \ .
\end{align}
We employ the Hatree-Fock (HF) approximation to the retarded SPGF:
\begin{align}
  \label{eq:hfgret}
  \vec{G}^\mathrm{R}(\vec{k};t,t^\prime) &=
  -\iu\theta(t-t^\prime)\mathcal{T} \exp\left(\int^t_{t^\prime}\!\dd
  \bar{t}\, \vec{h}^{\mathrm{HF}}(\vec{k};\bar{t}) \right) \nonumber
  \\ &\equiv  -\iu\theta(t-t^\prime) \vec{U}(\vec{k};t,t^\prime) ,
\end{align}
where $\vec{h}^{\mathrm{HF}}(\vec{k};\bar{t})$ denotes the mean-field
HF Hamiltonian, while $\mathcal{T}$ stands for the time-ordering
symbol. The time-evolution operator $\vec{U}(\vec{k};t,t^\prime)$
defined by Eq.~\eqref{eq:hfgret} is computed using the semi-group
property
$\vec{U}(\vec{k};t_n+\Delta t,t_j) = \vec{U}(\vec{k};t_n+\Delta t,t_n)
\vec{U}(\vec{k};t_n,t_j)$ on a uniform mesh of time points
$t_n = n\Delta t$. The propagator $\vec{U}(\vec{k};t_n+\Delta t,t_n)$
is computed using the fourth-order commutator-free matrix-exponential
method~\cite{alvermann_high-order_2011}. The GKBA
equation~\eqref{eq:gkba1} is solved using an in-house highly accurate
computer code. A fixed time step $\Delta t=0.05$ was used in all calculations.

\subsection{Self-energy: second-Born approximation}

\begin{figure}[t]
  \includegraphics[width=\columnwidth]{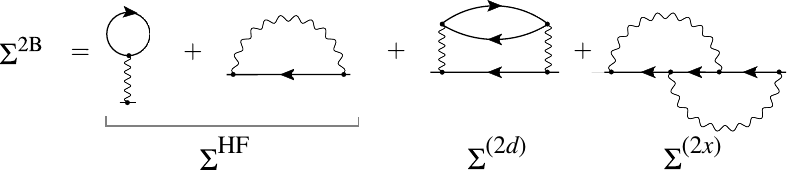}
  \caption{Feynmann diagrams representing the second-Born
    approximation, consisting of the Hatree-Fock (first two diagrams),
    direct (third diagram) and exchange (last diagram) contribution.\label{fig:2b}}
\end{figure}

All results in the main text have been obtained within the second-Born
approximation (2BA). The corresponding diagrammatic representation is
shown in Fig.~\ref{fig:2b}.

Using the standard Feynmann
rules~\cite{stefanucci_nonequilibrium_2013}, the diagrams for the 2BA have been
cast into mathematical expressions on the Kadanoff-Baym contour and
implemented in our computer codes. A general explicit expression in the Wannier
representation can be found, for instance, in 
Ref.~\cite{schlunzen_nonequilibrium_2016}. 

For nonlocal interactions, however, the large computational effort to
treat the exchange diagram prevents us from employing the full 2BA in this
case. Therefore, we have omitted $\Sigma^{(2x)}$ in the treatment of
the Chern insulators with nonlocal interactions. For all other cases,
we have confirmed that not including the exchange diagrams leads to
very small quantitative changes. Hence, all statements in the
main text on
thermalization still remain valid. Note that even without including
the exchange diagram, the resulting 2BA is still energy conserving.

\section{Appendix: Scaling of purity gap closing}

\begin{figure}[b]
  \includegraphics[width=0.9\columnwidth]{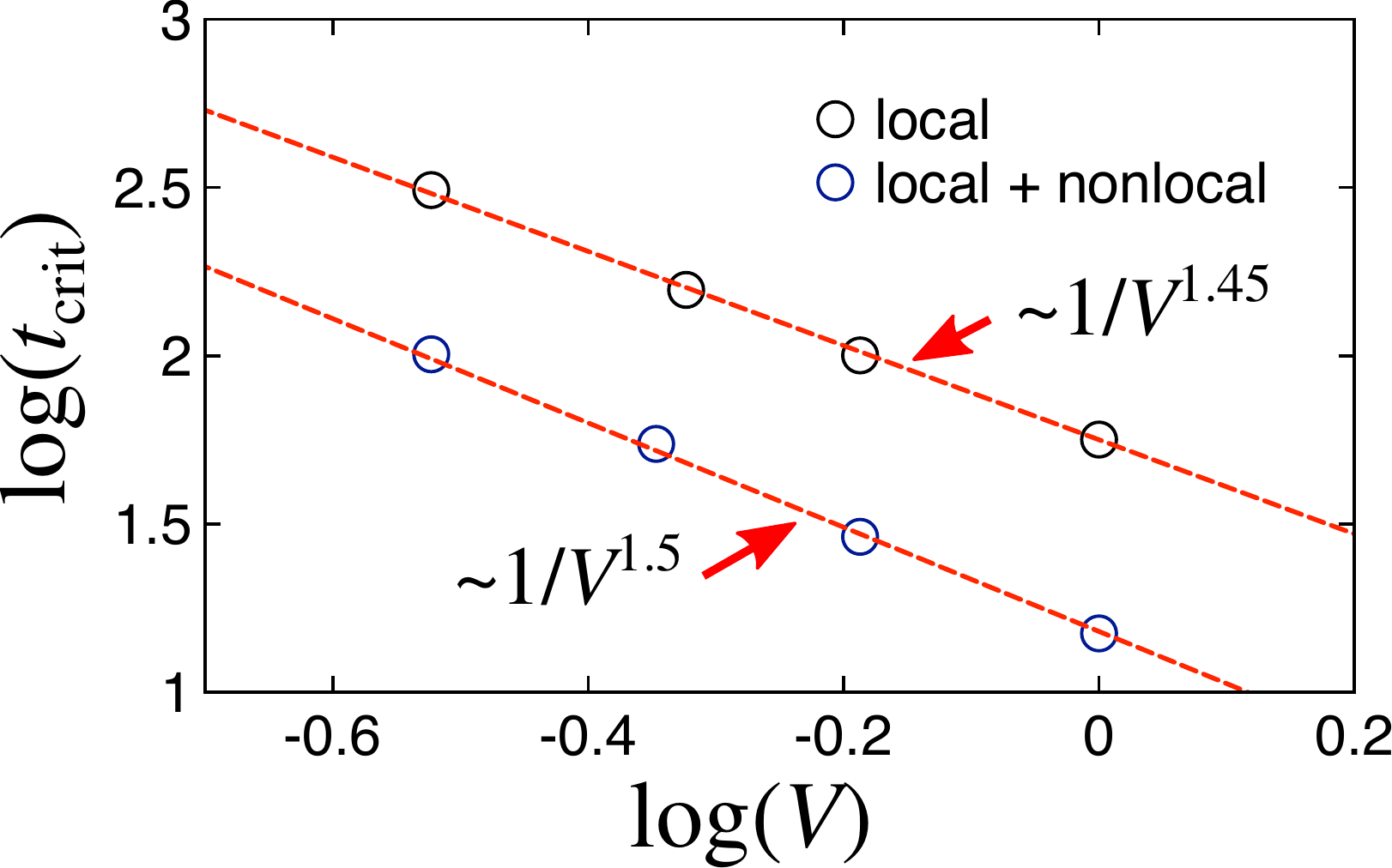}
  \caption{Double-logarithmic plot of the critical time of the purity
    gap closing as function of the interaction strength.\label{fig:tcrit}}
\end{figure}

In the main text, we have discussed the purity gap closing
characterized by the critical time $t_\mathrm{crit}$. In order to
investigate the dependence on the interaction, we have computed
$t_\mathrm{crit}$ for additional values of the interaction strength
for both the Chern insulator with local interactions only and
including nonlocal interactions. The result is presented in
Fig.~\ref{fig:tcrit}.

Linear regression of $\log(t_\mathrm{crit})$ as a function of $\log(V)$
shows that in the scenario with or without nonlocal interactions the
critical time approximately scales as $\sim V^{-3/2}$. Interstingly,
long-time relaxation times scale as $\sim V^{-2}$; 
hence, the time
scale of the purity gap closing is different from thermalization and
more related to dephasing effects.

\section{Appendix: Calculation of the equilibrium Hall conductance}

\begin{figure}[t]
  \includegraphics[width=\columnwidth]{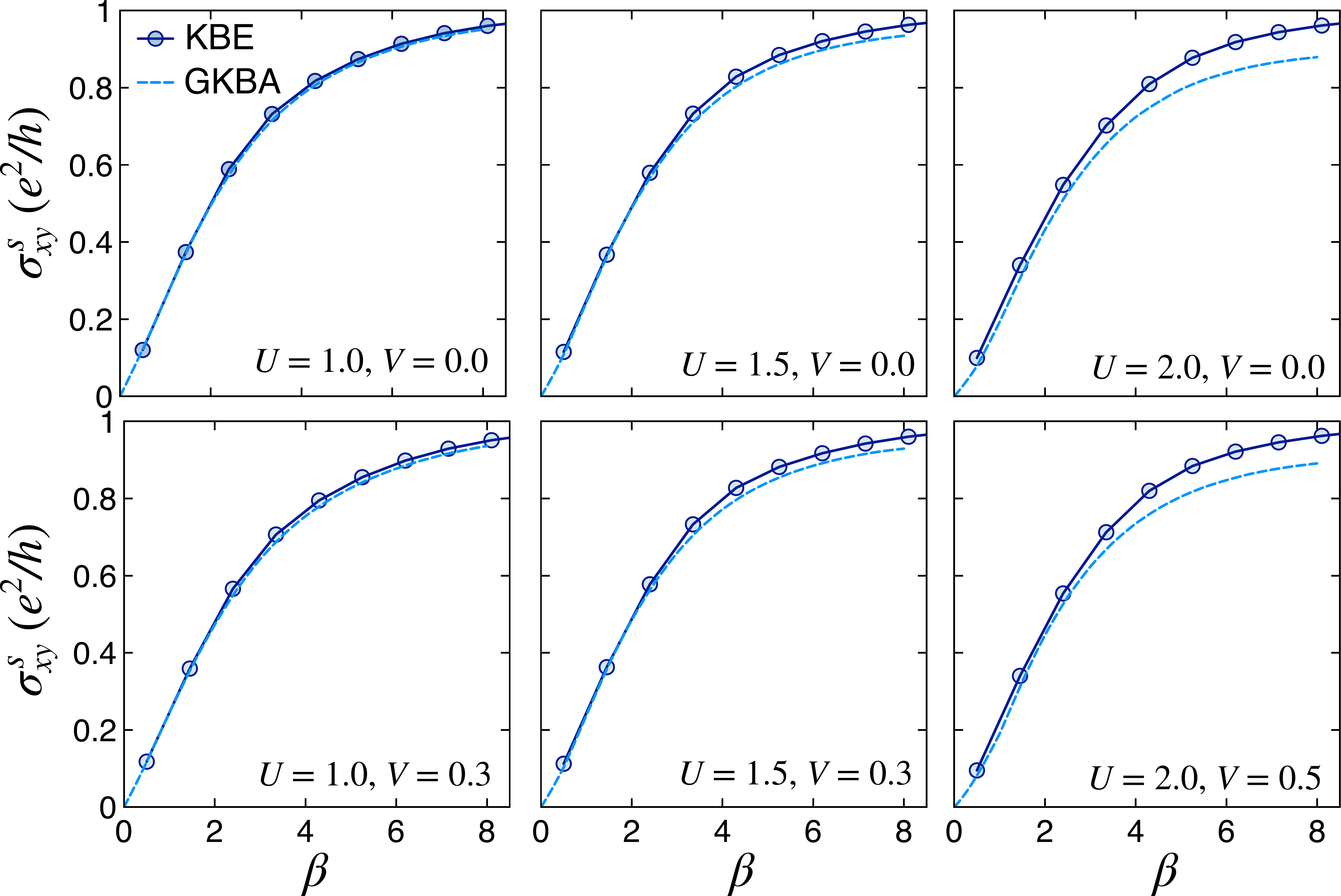}
  \caption{Hall conductance of the spinfull $\mathbb{Z}_2$ insulator
    in thermal equilibrium as a function of the inverse temperature
    $\beta$ for different values of the local interaction parameters
    $U$ and $V$. \label{fig:eqhall}}
\end{figure}

In order to investigate if the steady-state Hall conductance discussed
in the main text corresponds to thermal equilibrium, we computed the
equilibrium Hall conductance as a function of temperature. Following
Ref.~\cite{latini_charge_2014}, we have prepared the initial density
matrix $\gvec{\rho}_\sigma(\vec{k},t=0)$ with respect to the
topologically nontrival post-quench Hamiltonian including interactions
on the mean-field level. Propagating using the GKBA while adiabatically
switching on the 2BA self-energy yields a correlated initial state
$\gvec{\rho}_\sigma(\vec{k},t_\mathrm{switch})$. Applying the probe electric
field $E_y(t)= F_0(1-e^{-(t-t_\mathrm{switch})/\tau_0})$ ($E_y(t) = 0$
for $t<t_\mathrm{switch}$) after the interactions are switched on then
yields the equilibrium Hall conductance via $\sigma_{xy} =
\lim_{t\rightarrow\infty} J_x(t)/F_0$. This procedure is performed for
a set of inverse  temperatures $\beta$. Repeating the adiabatic switching procedure
without probe field leads to a constant total energy
$E_\mathrm{tot}$, which yields the temperature dependence of
$E_\mathrm{tot}$. The function $E_\mathrm{tot}(\beta)$ is then used to
determine the effective temperature $T_\mathrm{eff}$.  The adiabatic
switching was realized using the double-exponential switch-on function
from ref.~\cite{schlunzen_nonequilibrium_2016}, using a time interval
of $t_\mathrm{switch}=40$. 

Within the full KBE treatment, on the other hand, the preparation of a
correlated initial state in thermal equilibrium is accomplished by
solving the Dyson equation for the Matsubara SPGF
$\vec{G}^\mathrm{M}(\vec{k};\tau)$. The total energy
$E_\mathrm{tot}(\beta)$ is computed via the Galitskii-Migdal
formula~\cite{stefanucci_nonequilibrium_2013}. The time evolution
in the presence of the probe field $E_y(t)= F_0(1-e^{-(t/\tau_0})$ is then
obtained by solving the full set of the KBEs~\eqref{eq:kbes}.

Figure~\ref{fig:eqhall} shows the spin Hall conductance of the
$\mathbb{Z}_2$ insulator in thermal equilibrium, comparing the full
KBE and the GKBA treatment. The agreement is very good for smaller
$\beta$ (higher temperature, that is) and weaker interactions, while
deviations become apparent for low temperature and stronger
interaction. In particular, the full KBE treatment recovers the limit
$\sigma^s_{xy} \rightarrow e^2/h$ for $\beta \rightarrow \infty$. 
This is consistent with the fact that the topological properties can
not be altered by (weak) electron-electron interactions. In contrast,
the GKBA does not reproduce this limit correctly. Nevertheless, since
the effective temperatures in the quench setup studied in the main
text are quite high (typically $\beta\sim 1$ to $\beta\sim 2$), the
GKBA provides an accurate description.

\section{Appendix: Dependence on pre- and post-quench gap parameter}

\begin{figure}[t]
  \includegraphics[width=\columnwidth]{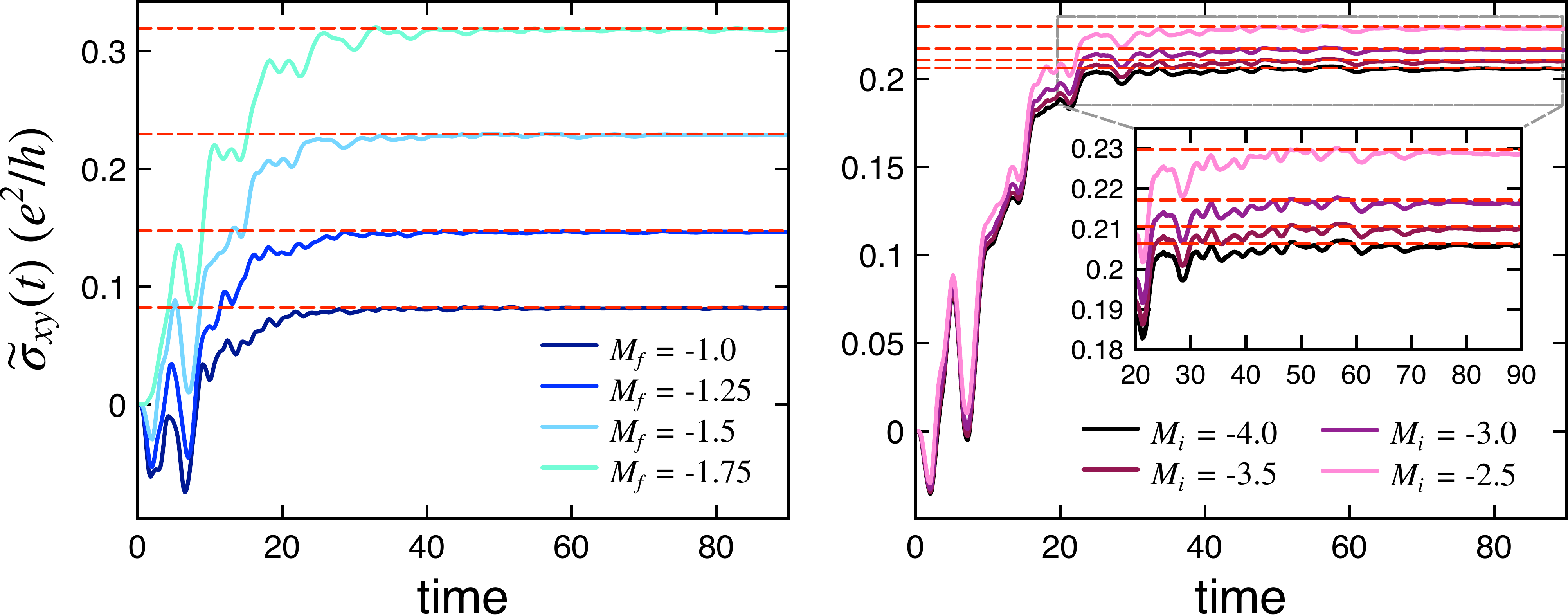}
  \caption{Nonequilibrium (spin) Hall conductance of the spinfull
    $\mathbb{Z}_2$ insulattor for the same setup
    as in the main text. In the left panel, $M_i=-3.5$, while
    $M_f=-1.5$ for the panel on the right-hand side. The red-dashed
    lines indicate the corresponding thermal equilibrium. \label{fig:gap}}
\end{figure}

Figure~\ref{fig:gap} shows the nonequilibrium Hall conductance
$\widetilde{\sigma}_{xy}(t)$ of the spin Chern insulator in analogy to the result in the main text
for different values of the pre- ($M_i$) and post-quench ($M_f$) mass
parameter. The parameters are the same as in the main text. The
Coulomb interaction has been fixed to $U=1.5, V=0.0$. We have used the
GKBA to obtain the results in Fig.~\ref{fig:gap}.

As Fig.~\ref{fig:gap} demonstrates, the nonequilibrium Hall
conductance approaches the respective thermal equilibrium value for
all parameters. Inspecting the dependence on the initial mass
parameter $M_i$ (right-hand panel in Fig.~\ref{fig:gap}), one finds a
quite weak variation. This indicates that the gap size of the
pre-quench band insulator plays only a minor role, since the band
hybridization can be neglected. In contrast, the dependence on the
post-quench gap parameter $M_f$ (left-hand panel in
Fig.~\ref{fig:gap}) is much more pronounced. One finds an increasing
Hall conductance with $M_f$ approaching the phase boundary
$M_f=-2$. This can be understood by the amount of energy injected by
the quench, which is minimal for $M_f\rightarrow -2$. Therefore, the
steady-state Hall conductance is the largest for $M_f = -1.75$. This
dependence explains the dependence of $\sigma^s_{xy}$ of the spin
Chern insulator on $U$ discussed in the main text: right after quench,
the effective mean-field Hamiltonian $\vec{h}^\mathrm{HF}_{\uparrow,\downarrow}(\vec{k})$
becomes
\begin{align}
  \vec{h}^\mathrm{HF}_{\uparrow,\downarrow} (\vec{k}) =
  \vec{h}_{\uparrow,\downarrow} (\vec{k}) + U 
  \begin{pmatrix} n_E + \frac12 & 0 \\ 0 & n_E - \frac12 \end{pmatrix}
                                           \ .
\end{align}
Since the lower band of the pre-quench band insulator is predominantly
consisting of the $E$-orbital, the occupation $n_E \approx 1$, thus
$\vec{h}^\mathrm{HF}_{\uparrow,\downarrow} (\vec{k}) \approx
\vec{h}_{\uparrow,\downarrow} (\vec{k}) +\frac{U}{2}
\sigma_z$, corresponding to the free Hamiltonian
$\vec{h}_{\uparrow,\downarrow} (\vec{k})$ with $M\rightarrow M -
U/2$. Hence, increasing $U$ results in an effective post-quench
mass parameter deeper in the topological phase, and thus a larger energy injection
and a reduced steady-state Hall conductance.


\begin{thebibliography}{57}%
\makeatletter
\providecommand \@ifxundefined [1]{%
 \@ifx{#1\undefined}
}%
\providecommand \@ifnum [1]{%
 \ifnum #1\expandafter \@firstoftwo
 \else \expandafter \@secondoftwo
 \fi
}%
\providecommand \@ifx [1]{%
 \ifx #1\expandafter \@firstoftwo
 \else \expandafter \@secondoftwo
 \fi
}%
\providecommand \natexlab [1]{#1}%
\providecommand \enquote  [1]{``#1''}%
\providecommand \bibnamefont  [1]{#1}%
\providecommand \bibfnamefont [1]{#1}%
\providecommand \citenamefont [1]{#1}%
\providecommand \href@noop [0]{\@secondoftwo}%
\providecommand \href [0]{\begingroup \@sanitize@url \@href}%
\providecommand \@href[1]{\@@startlink{#1}\@@href}%
\providecommand \@@href[1]{\endgroup#1\@@endlink}%
\providecommand \@sanitize@url [0]{\catcode `\\12\catcode `\$12\catcode
  `\&12\catcode `\#12\catcode `\^12\catcode `\_12\catcode `\%12\relax}%
\providecommand \@@startlink[1]{}%
\providecommand \@@endlink[0]{}%
\providecommand \url  [0]{\begingroup\@sanitize@url \@url }%
\providecommand \@url [1]{\endgroup\@href {#1}{\urlprefix }}%
\providecommand \urlprefix  [0]{URL }%
\providecommand \Eprint [0]{\href }%
\providecommand \doibase [0]{http://dx.doi.org/}%
\providecommand \selectlanguage [0]{\@gobble}%
\providecommand \bibinfo  [0]{\@secondoftwo}%
\providecommand \bibfield  [0]{\@secondoftwo}%
\providecommand \translation [1]{[#1]}%
\providecommand \BibitemOpen [0]{}%
\providecommand \bibitemStop [0]{}%
\providecommand \bibitemNoStop [0]{.\EOS\space}%
\providecommand \EOS [0]{\spacefactor3000\relax}%
\providecommand \BibitemShut  [1]{\csname bibitem#1\endcsname}%
\let\auto@bib@innerbib\@empty
\bibitem [{\citenamefont {{Aidelsburger}}\ \emph {et~al.}(2013)\citenamefont
  {{Aidelsburger}}, \citenamefont {{Atala}}, \citenamefont {{Lohse}},
  \citenamefont {{Barreiro}}, \citenamefont {{Paredes}},\ and\ \citenamefont
  {{Bloch}}}]{Aidelsburger2013}%
  \BibitemOpen
  \bibfield  {author} {\bibinfo {author} {\bibfnamefont {M.}~\bibnamefont
  {{Aidelsburger}}}, \bibinfo {author} {\bibfnamefont {M.}~\bibnamefont
  {{Atala}}}, \bibinfo {author} {\bibfnamefont {M.}~\bibnamefont {{Lohse}}},
  \bibinfo {author} {\bibfnamefont {J.~T.}\ \bibnamefont {{Barreiro}}},
  \bibinfo {author} {\bibfnamefont {B.}~\bibnamefont {{Paredes}}}, \ and\
  \bibinfo {author} {\bibfnamefont {I.}~\bibnamefont {{Bloch}}},\ }\href@noop
  {} {\bibfield  {journal} {\bibinfo  {journal} {\prl}\ }\textbf {\bibinfo
  {volume} {111}},\ \bibinfo {eid} {185301} (\bibinfo {year}
  {2013})}\BibitemShut {NoStop}%
\bibitem [{\citenamefont {{Miyake}}\ \emph {et~al.}(2013)\citenamefont
  {{Miyake}}, \citenamefont {{Siviloglou}}, \citenamefont {{Kennedy}},
  \citenamefont {{Burton}},\ and\ \citenamefont {{Ketterle}}}]{Ketterle2013}%
  \BibitemOpen
  \bibfield  {author} {\bibinfo {author} {\bibfnamefont {H.}~\bibnamefont
  {{Miyake}}}, \bibinfo {author} {\bibfnamefont {G.~A.}\ \bibnamefont
  {{Siviloglou}}}, \bibinfo {author} {\bibfnamefont {C.~J.}\ \bibnamefont
  {{Kennedy}}}, \bibinfo {author} {\bibfnamefont {W.~C.}\ \bibnamefont
  {{Burton}}}, \ and\ \bibinfo {author} {\bibfnamefont {W.}~\bibnamefont
  {{Ketterle}}},\ }\href@noop {} {\bibfield  {journal} {\bibinfo  {journal}
  {\prl}\ }\textbf {\bibinfo {volume} {111}},\ \bibinfo {eid} {185302}
  (\bibinfo {year} {2013})}\BibitemShut {NoStop}%
\bibitem [{\citenamefont {Jotzu}\ \emph {et~al.}(2014)\citenamefont {Jotzu},
  \citenamefont {Messer}, \citenamefont {Desbuquois}, \citenamefont {Lebrat},
  \citenamefont {Uehlinger}, \citenamefont {Greif},\ and\ \citenamefont
  {Esslinger}}]{jotzu_experimental_2014}%
  \BibitemOpen
  \bibfield  {author} {\bibinfo {author} {\bibfnamefont {G.}~\bibnamefont
  {Jotzu}}, \bibinfo {author} {\bibfnamefont {M.}~\bibnamefont {Messer}},
  \bibinfo {author} {\bibfnamefont {R.}~\bibnamefont {Desbuquois}}, \bibinfo
  {author} {\bibfnamefont {M.}~\bibnamefont {Lebrat}}, \bibinfo {author}
  {\bibfnamefont {T.}~\bibnamefont {Uehlinger}}, \bibinfo {author}
  {\bibfnamefont {D.}~\bibnamefont {Greif}}, \ and\ \bibinfo {author}
  {\bibfnamefont {T.}~\bibnamefont {Esslinger}},\ }\href@noop {} {\bibfield
  {journal} {\bibinfo  {journal} {Nature}\ }\textbf {\bibinfo {volume} {515}},\
  \bibinfo {pages} {237} (\bibinfo {year} {2014})}\BibitemShut {NoStop}%
\bibitem [{\citenamefont {Wu}\ \emph {et~al.}(2016)\citenamefont {Wu},
  \citenamefont {Zhang}, \citenamefont {Sun}, \citenamefont {Xu}, \citenamefont
  {Wang}, \citenamefont {Ji}, \citenamefont {Deng}, \citenamefont {Chen},
  \citenamefont {Liu},\ and\ \citenamefont {Pan}}]{wu_realization_2016-1}%
  \BibitemOpen
  \bibfield  {author} {\bibinfo {author} {\bibfnamefont {Z.}~\bibnamefont
  {Wu}}, \bibinfo {author} {\bibfnamefont {L.}~\bibnamefont {Zhang}}, \bibinfo
  {author} {\bibfnamefont {W.}~\bibnamefont {Sun}}, \bibinfo {author}
  {\bibfnamefont {X.-T.}\ \bibnamefont {Xu}}, \bibinfo {author} {\bibfnamefont
  {B.-Z.}\ \bibnamefont {Wang}}, \bibinfo {author} {\bibfnamefont {S.-C.}\
  \bibnamefont {Ji}}, \bibinfo {author} {\bibfnamefont {Y.}~\bibnamefont
  {Deng}}, \bibinfo {author} {\bibfnamefont {S.}~\bibnamefont {Chen}}, \bibinfo
  {author} {\bibfnamefont {X.-J.}\ \bibnamefont {Liu}}, \ and\ \bibinfo
  {author} {\bibfnamefont {J.-W.}\ \bibnamefont {Pan}},\ }\href@noop {}
  {\bibfield  {journal} {\bibinfo  {journal} {Science}\ }\textbf {\bibinfo
  {volume} {354}},\ \bibinfo {pages} {83} (\bibinfo {year} {2016})}\BibitemShut
  {NoStop}%
\bibitem [{\citenamefont {Fl\"{a}schner}\ \emph {et~al.}(2016)\citenamefont
  {Fl\"{a}schner}, \citenamefont {Rem}, \citenamefont {Tarnowski},
  \citenamefont {Vogel}, \citenamefont {L\"{u}hmann}, \citenamefont
  {Sengstock},\ and\ \citenamefont {Weitenberg}}]{flaschner_experimental_2016}%
  \BibitemOpen
  \bibfield  {author} {\bibinfo {author} {\bibfnamefont {N.}~\bibnamefont
  {Fl\"{a}schner}}, \bibinfo {author} {\bibfnamefont {B.~S.}\ \bibnamefont
  {Rem}}, \bibinfo {author} {\bibfnamefont {M.}~\bibnamefont {Tarnowski}},
  \bibinfo {author} {\bibfnamefont {D.}~\bibnamefont {Vogel}}, \bibinfo
  {author} {\bibfnamefont {D.-S.}\ \bibnamefont {L\"{u}hmann}}, \bibinfo
  {author} {\bibfnamefont {K.}~\bibnamefont {Sengstock}}, \ and\ \bibinfo
  {author} {\bibfnamefont {C.}~\bibnamefont {Weitenberg}},\ }\href@noop {}
  {\bibfield  {journal} {\bibinfo  {journal} {Science}\ }\textbf {\bibinfo
  {volume} {352}},\ \bibinfo {pages} {1091} (\bibinfo {year}
  {2016})}\BibitemShut {NoStop}%
\bibitem [{\citenamefont {{Goldman}}\ \emph {et~al.}(2016)\citenamefont
  {{Goldman}}, \citenamefont {{Budich}},\ and\ \citenamefont
  {{Zoller}}}]{GoldmanReview2016}%
  \BibitemOpen
  \bibfield  {author} {\bibinfo {author} {\bibfnamefont {N.}~\bibnamefont
  {{Goldman}}}, \bibinfo {author} {\bibfnamefont {J.~C.}\ \bibnamefont
  {{Budich}}}, \ and\ \bibinfo {author} {\bibfnamefont {P.}~\bibnamefont
  {{Zoller}}},\ }\href@noop {} {\bibfield  {journal} {\bibinfo  {journal}
  {Nature Physics}\ }\textbf {\bibinfo {volume} {12}},\ \bibinfo {pages} {639}
  (\bibinfo {year} {2016})}\BibitemShut {NoStop}%
\bibitem [{\citenamefont {Wang}\ \emph {et~al.}(2013)\citenamefont {Wang},
  \citenamefont {Steinberg}, \citenamefont {Jarillo-Herrero},\ and\
  \citenamefont {Gedik}}]{wang_observation_2013}%
  \BibitemOpen
  \bibfield  {author} {\bibinfo {author} {\bibfnamefont {Y.~H.}\ \bibnamefont
  {Wang}}, \bibinfo {author} {\bibfnamefont {H.}~\bibnamefont {Steinberg}},
  \bibinfo {author} {\bibfnamefont {P.}~\bibnamefont {Jarillo-Herrero}}, \ and\
  \bibinfo {author} {\bibfnamefont {N.}~\bibnamefont {Gedik}},\ }\href@noop {}
  {\bibfield  {journal} {\bibinfo  {journal} {Science}\ }\textbf {\bibinfo
  {volume} {342}},\ \bibinfo {pages} {453} (\bibinfo {year}
  {2013})}\BibitemShut {NoStop}%
\bibitem [{\citenamefont {McIver}\ \emph {et~al.}(2018)\citenamefont {McIver},
  \citenamefont {Schulte}, \citenamefont {Stein}, \citenamefont {Matsuyama},
  \citenamefont {Jotzu}, \citenamefont {Meier},\ and\ \citenamefont
  {Cavalleri}}]{mciver_light-induced_2018}%
  \BibitemOpen
  \bibfield  {author} {\bibinfo {author} {\bibfnamefont {J.~W.}\ \bibnamefont
  {McIver}}, \bibinfo {author} {\bibfnamefont {B.}~\bibnamefont {Schulte}},
  \bibinfo {author} {\bibfnamefont {F.-U.}\ \bibnamefont {Stein}}, \bibinfo
  {author} {\bibfnamefont {T.}~\bibnamefont {Matsuyama}}, \bibinfo {author}
  {\bibfnamefont {G.}~\bibnamefont {Jotzu}}, \bibinfo {author} {\bibfnamefont
  {G.}~\bibnamefont {Meier}}, \ and\ \bibinfo {author} {\bibfnamefont
  {A.}~\bibnamefont {Cavalleri}},\ }\href@noop {} {\bibfield  {journal}
  {\bibinfo  {journal} {arXiv:1811.03522 [cond-mat]}\ } (\bibinfo {year}
  {2018})}\BibitemShut {NoStop}%
\bibitem [{\citenamefont {{Chen}}\ \emph {et~al.}(2010)\citenamefont {{Chen}},
  \citenamefont {{Gu}},\ and\ \citenamefont {{Wen}}}]{ChenLUT2010}%
  \BibitemOpen
  \bibfield  {author} {\bibinfo {author} {\bibfnamefont {X.}~\bibnamefont
  {{Chen}}}, \bibinfo {author} {\bibfnamefont {Z.-C.}\ \bibnamefont {{Gu}}}, \
  and\ \bibinfo {author} {\bibfnamefont {X.-G.}\ \bibnamefont {{Wen}}},\
  }\href@noop {} {\bibfield  {journal} {\bibinfo  {journal} {Physical Review
  B}\ }\textbf {\bibinfo {volume} {82}},\ \bibinfo {eid} {155138} (\bibinfo
  {year} {2010})}\BibitemShut {NoStop}%
\bibitem [{\citenamefont {Tran}\ \emph {et~al.}(2017)\citenamefont {Tran},
  \citenamefont {Dauphin}, \citenamefont {Grushin}, \citenamefont {Zoller},\
  and\ \citenamefont {Goldman}}]{tran_probing_2017-1}%
  \BibitemOpen
  \bibfield  {author} {\bibinfo {author} {\bibfnamefont {D.~T.}\ \bibnamefont
  {Tran}}, \bibinfo {author} {\bibfnamefont {A.}~\bibnamefont {Dauphin}},
  \bibinfo {author} {\bibfnamefont {A.~G.}\ \bibnamefont {Grushin}}, \bibinfo
  {author} {\bibfnamefont {P.}~\bibnamefont {Zoller}}, \ and\ \bibinfo {author}
  {\bibfnamefont {N.}~\bibnamefont {Goldman}},\ }\href@noop {} {\bibfield
  {journal} {\bibinfo  {journal} {Science Advances}\ }\textbf {\bibinfo
  {volume} {3}},\ \bibinfo {pages} {e1701207} (\bibinfo {year}
  {2017})}\BibitemShut {NoStop}%
\bibitem [{\citenamefont {Sch\"{u}ler}\ and\ \citenamefont
  {Werner}(2017)}]{schuler_tracing_2017}%
  \BibitemOpen
  \bibfield  {author} {\bibinfo {author} {\bibfnamefont {M.}~\bibnamefont
  {Sch\"{u}ler}}\ and\ \bibinfo {author} {\bibfnamefont {P.}~\bibnamefont
  {Werner}},\ }\href@noop {} {\bibfield  {journal} {\bibinfo  {journal} {Phys.
  Rev. B}\ }\textbf {\bibinfo {volume} {96}},\ \bibinfo {pages} {155122}
  (\bibinfo {year} {2017})}\BibitemShut {NoStop}%
\bibitem [{\citenamefont {Tran}\ \emph {et~al.}(2018)\citenamefont {Tran},
  \citenamefont {Cooper},\ and\ \citenamefont {Goldman}}]{tran_quantized_2018}%
  \BibitemOpen
  \bibfield  {author} {\bibinfo {author} {\bibfnamefont {D.~T.}\ \bibnamefont
  {Tran}}, \bibinfo {author} {\bibfnamefont {N.~R.}\ \bibnamefont {Cooper}}, \
  and\ \bibinfo {author} {\bibfnamefont {N.}~\bibnamefont {Goldman}},\
  }\href@noop {} {\bibfield  {journal} {\bibinfo  {journal} {Phys. Rev. A}\
  }\textbf {\bibinfo {volume} {97}},\ \bibinfo {pages} {061602} (\bibinfo
  {year} {2018})}\BibitemShut {NoStop}%
\bibitem [{\citenamefont {Caio}\ \emph {et~al.}(2015)\citenamefont {Caio},
  \citenamefont {Cooper},\ and\ \citenamefont {Bhaseen}}]{caio_quantum_2015}%
  \BibitemOpen
  \bibfield  {author} {\bibinfo {author} {\bibfnamefont {M.}~\bibnamefont
  {Caio}}, \bibinfo {author} {\bibfnamefont {N.}~\bibnamefont {Cooper}}, \ and\
  \bibinfo {author} {\bibfnamefont {M.}~\bibnamefont {Bhaseen}},\ }\href@noop
  {} {\bibfield  {journal} {\bibinfo  {journal} {Phys. Rev. Lett.}\ }\textbf
  {\bibinfo {volume} {115}},\ \bibinfo {pages} {236403} (\bibinfo {year}
  {2015})}\BibitemShut {NoStop}%
\bibitem [{\citenamefont {Dehghani}\ and\ \citenamefont
  {Mitra}(2015)}]{dehghani_optical_2015}%
  \BibitemOpen
  \bibfield  {author} {\bibinfo {author} {\bibfnamefont {H.}~\bibnamefont
  {Dehghani}}\ and\ \bibinfo {author} {\bibfnamefont {A.}~\bibnamefont
  {Mitra}},\ }\href@noop {} {\bibfield  {journal} {\bibinfo  {journal} {Phys.
  Rev. B}\ }\textbf {\bibinfo {volume} {92}},\ \bibinfo {pages} {165111}
  (\bibinfo {year} {2015})}\BibitemShut {NoStop}%
\bibitem [{\citenamefont {Dehghani}\ \emph {et~al.}(2015)\citenamefont
  {Dehghani}, \citenamefont {Oka},\ and\ \citenamefont
  {Mitra}}]{dehghani_out--equilibrium_2015}%
  \BibitemOpen
  \bibfield  {author} {\bibinfo {author} {\bibfnamefont {H.}~\bibnamefont
  {Dehghani}}, \bibinfo {author} {\bibfnamefont {T.}~\bibnamefont {Oka}}, \
  and\ \bibinfo {author} {\bibfnamefont {A.}~\bibnamefont {Mitra}},\
  }\href@noop {} {\bibfield  {journal} {\bibinfo  {journal} {Phys. Rev. B}\
  }\textbf {\bibinfo {volume} {91}},\ \bibinfo {pages} {155422} (\bibinfo
  {year} {2015})}\BibitemShut {NoStop}%
\bibitem [{\citenamefont {Wang}\ \emph {et~al.}(2016)\citenamefont {Wang},
  \citenamefont {Schmitt},\ and\ \citenamefont
  {Kehrein}}]{wang_universal_2016}%
  \BibitemOpen
  \bibfield  {author} {\bibinfo {author} {\bibfnamefont {P.}~\bibnamefont
  {Wang}}, \bibinfo {author} {\bibfnamefont {M.}~\bibnamefont {Schmitt}}, \
  and\ \bibinfo {author} {\bibfnamefont {S.}~\bibnamefont {Kehrein}},\
  }\href@noop {} {\bibfield  {journal} {\bibinfo  {journal} {Phys. Rev. B}\
  }\textbf {\bibinfo {volume} {93}},\ \bibinfo {pages} {085134} (\bibinfo
  {year} {2016})}\BibitemShut {NoStop}%
\bibitem [{\citenamefont {Hu}\ \emph {et~al.}(2016)\citenamefont {Hu},
  \citenamefont {Zoller},\ and\ \citenamefont {Budich}}]{hu_dynamical_2016}%
  \BibitemOpen
  \bibfield  {author} {\bibinfo {author} {\bibfnamefont {Y.}~\bibnamefont
  {Hu}}, \bibinfo {author} {\bibfnamefont {P.}~\bibnamefont {Zoller}}, \ and\
  \bibinfo {author} {\bibfnamefont {J.~C.}\ \bibnamefont {Budich}},\
  }\href@noop {} {\bibfield  {journal} {\bibinfo  {journal} {Phys. Rev. Lett.}\
  }\textbf {\bibinfo {volume} {117}},\ \bibinfo {pages} {126803} (\bibinfo
  {year} {2016})}\BibitemShut {NoStop}%
\bibitem [{\citenamefont {Schmitt}\ and\ \citenamefont
  {Wang}(2017)}]{schmitt_universal_2017-1}%
  \BibitemOpen
  \bibfield  {author} {\bibinfo {author} {\bibfnamefont {M.}~\bibnamefont
  {Schmitt}}\ and\ \bibinfo {author} {\bibfnamefont {P.}~\bibnamefont {Wang}},\
  }\href@noop {} {\bibfield  {journal} {\bibinfo  {journal} {Phys. Rev. B}\
  }\textbf {\bibinfo {volume} {96}},\ \bibinfo {pages} {054306} (\bibinfo
  {year} {2017})}\BibitemShut {NoStop}%
\bibitem [{\citenamefont {Ul\v{c}akar}\ \emph {et~al.}(2018)\citenamefont
  {Ul\v{c}akar}, \citenamefont {Mravlje}, \citenamefont {Ram\v{s}ak},\ and\
  \citenamefont {Rejec}}]{ulcakar_slow_2018}%
  \BibitemOpen
  \bibfield  {author} {\bibinfo {author} {\bibfnamefont {L.}~\bibnamefont
  {Ul\v{c}akar}}, \bibinfo {author} {\bibfnamefont {J.}~\bibnamefont
  {Mravlje}}, \bibinfo {author} {\bibfnamefont {A.}~\bibnamefont {Ram\v{s}ak}},
  \ and\ \bibinfo {author} {\bibfnamefont {T.}~\bibnamefont {Rejec}},\
  }\href@noop {} {\bibfield  {journal} {\bibinfo  {journal} {Phys. Rev. B}\
  }\textbf {\bibinfo {volume} {97}},\ \bibinfo {pages} {195127} (\bibinfo
  {year} {2018})}\BibitemShut {NoStop}%
\bibitem [{\citenamefont {Peralta~Gavensky}\ \emph {et~al.}(2018)\citenamefont
  {Peralta~Gavensky}, \citenamefont {Usaj},\ and\ \citenamefont
  {Balseiro}}]{peralta_gavensky_time-resolved_2018}%
  \BibitemOpen
  \bibfield  {author} {\bibinfo {author} {\bibfnamefont {L.}~\bibnamefont
  {Peralta~Gavensky}}, \bibinfo {author} {\bibfnamefont {G.}~\bibnamefont
  {Usaj}}, \ and\ \bibinfo {author} {\bibfnamefont {C.~A.}\ \bibnamefont
  {Balseiro}},\ }\href@noop {} {\bibfield  {journal} {\bibinfo  {journal}
  {Phys. Rev. B}\ }\textbf {\bibinfo {volume} {98}},\ \bibinfo {pages} {165414}
  (\bibinfo {year} {2018})}\BibitemShut {NoStop}%
\bibitem [{\citenamefont {Xu}\ and\ \citenamefont {Hu}(2018)}]{xu_scheme_2018}%
  \BibitemOpen
  \bibfield  {author} {\bibinfo {author} {\bibfnamefont {Y.}~\bibnamefont
  {Xu}}\ and\ \bibinfo {author} {\bibfnamefont {Y.}~\bibnamefont {Hu}},\
  }\href@noop {} {\bibfield  {journal} {\bibinfo  {journal} {arXiv:1807.09732
  [cond-mat, physics:quant-ph]}\ } (\bibinfo {year} {2018})}\BibitemShut
  {NoStop}%
\bibitem [{\citenamefont {Wolff}\ \emph {et~al.}(2016)\citenamefont {Wolff},
  \citenamefont {Sheikhan},\ and\ \citenamefont
  {Kollath}}]{wolff_dissipative_2016}%
  \BibitemOpen
  \bibfield  {author} {\bibinfo {author} {\bibfnamefont {S.}~\bibnamefont
  {Wolff}}, \bibinfo {author} {\bibfnamefont {A.}~\bibnamefont {Sheikhan}}, \
  and\ \bibinfo {author} {\bibfnamefont {C.}~\bibnamefont {Kollath}},\
  }\href@noop {} {\bibfield  {journal} {\bibinfo  {journal} {Phys. Rev. A}\
  }\textbf {\bibinfo {volume} {94}},\ \bibinfo {pages} {043609} (\bibinfo
  {year} {2016})}\BibitemShut {NoStop}%
\bibitem [{\citenamefont {Berges}\ \emph {et~al.}(2004)\citenamefont {Berges},
  \citenamefont {Bors\'{a}nyi},\ and\ \citenamefont
  {Wetterich}}]{berges_prethermalization_2004}%
  \BibitemOpen
  \bibfield  {author} {\bibinfo {author} {\bibfnamefont {J.}~\bibnamefont
  {Berges}}, \bibinfo {author} {\bibfnamefont {S.}~\bibnamefont
  {Bors\'{a}nyi}}, \ and\ \bibinfo {author} {\bibfnamefont {C.}~\bibnamefont
  {Wetterich}},\ }\href@noop {} {\bibfield  {journal} {\bibinfo  {journal}
  {Phys. Rev. Lett.}\ }\textbf {\bibinfo {volume} {93}},\ \bibinfo {pages}
  {142002} (\bibinfo {year} {2004})}\BibitemShut {NoStop}%
\bibitem [{\citenamefont {Moeckel}\ and\ \citenamefont
  {Kehrein}(2008)}]{moeckel_interaction_2008}%
  \BibitemOpen
  \bibfield  {author} {\bibinfo {author} {\bibfnamefont {M.}~\bibnamefont
  {Moeckel}}\ and\ \bibinfo {author} {\bibfnamefont {S.}~\bibnamefont
  {Kehrein}},\ }\href@noop {} {\bibfield  {journal} {\bibinfo  {journal} {Phys.
  Rev. Lett.}\ }\textbf {\bibinfo {volume} {100}},\ \bibinfo {pages} {175702}
  (\bibinfo {year} {2008})}\BibitemShut {NoStop}%
\bibitem [{\citenamefont {Eckstein}\ \emph {et~al.}(2009)\citenamefont
  {Eckstein}, \citenamefont {Kollar},\ and\ \citenamefont
  {Werner}}]{eckstein_thermalization_2009}%
  \BibitemOpen
  \bibfield  {author} {\bibinfo {author} {\bibfnamefont {M.}~\bibnamefont
  {Eckstein}}, \bibinfo {author} {\bibfnamefont {M.}~\bibnamefont {Kollar}}, \
  and\ \bibinfo {author} {\bibfnamefont {P.}~\bibnamefont {Werner}},\
  }\href@noop {} {\bibfield  {journal} {\bibinfo  {journal} {Phys. Rev. Lett.}\
  }\textbf {\bibinfo {volume} {103}},\ \bibinfo {pages} {056403} (\bibinfo
  {year} {2009})}\BibitemShut {NoStop}%
\bibitem [{\citenamefont {Marcuzzi}\ \emph {et~al.}(2013)\citenamefont
  {Marcuzzi}, \citenamefont {Marino}, \citenamefont {Gambassi},\ and\
  \citenamefont {Silva}}]{marcuzzi_prethermalization_2013}%
  \BibitemOpen
  \bibfield  {author} {\bibinfo {author} {\bibfnamefont {M.}~\bibnamefont
  {Marcuzzi}}, \bibinfo {author} {\bibfnamefont {J.}~\bibnamefont {Marino}},
  \bibinfo {author} {\bibfnamefont {A.}~\bibnamefont {Gambassi}}, \ and\
  \bibinfo {author} {\bibfnamefont {A.}~\bibnamefont {Silva}},\ }\href@noop {}
  {\bibfield  {journal} {\bibinfo  {journal} {Phys. Rev. Lett.}\ }\textbf
  {\bibinfo {volume} {111}},\ \bibinfo {pages} {197203} (\bibinfo {year}
  {2013})}\BibitemShut {NoStop}%
\bibitem [{\citenamefont {D'Alessio}\ \emph {et~al.}(2016)\citenamefont
  {D'Alessio}, \citenamefont {Kafri}, \citenamefont {Polkovnikov},\ and\
  \citenamefont {Rigol}}]{dalessio_quantum_2016}%
  \BibitemOpen
  \bibfield  {author} {\bibinfo {author} {\bibfnamefont {L.}~\bibnamefont
  {D'Alessio}}, \bibinfo {author} {\bibfnamefont {Y.}~\bibnamefont {Kafri}},
  \bibinfo {author} {\bibfnamefont {A.}~\bibnamefont {Polkovnikov}}, \ and\
  \bibinfo {author} {\bibfnamefont {M.}~\bibnamefont {Rigol}},\ }\href@noop {}
  {\bibfield  {journal} {\bibinfo  {journal} {Advances in Physics}\ }\textbf
  {\bibinfo {volume} {65}},\ \bibinfo {pages} {239} (\bibinfo {year}
  {2016})}\BibitemShut {NoStop}%
\bibitem [{\citenamefont {Kruckenhauser}\ and\ \citenamefont
  {Budich}(2017)}]{kruckenhauser_dynamical_2017}%
  \BibitemOpen
  \bibfield  {author} {\bibinfo {author} {\bibfnamefont {A.}~\bibnamefont
  {Kruckenhauser}}\ and\ \bibinfo {author} {\bibfnamefont {J.~C.}\ \bibnamefont
  {Budich}},\ }\href@noop {} {\bibfield  {journal} {\bibinfo  {journal}
  {arXiv:1712.02440 [cond-mat, physics:quant-ph]}\ } (\bibinfo {year}
  {2017})}\BibitemShut {NoStop}%
\bibitem [{\citenamefont {Stan}\ \emph {et~al.}(2009)\citenamefont {Stan},
  \citenamefont {Dahlen},\ and\ \citenamefont {Leeuwen}}]{stan_time_2009}%
  \BibitemOpen
  \bibfield  {author} {\bibinfo {author} {\bibfnamefont {A.}~\bibnamefont
  {Stan}}, \bibinfo {author} {\bibfnamefont {N.~E.}\ \bibnamefont {Dahlen}}, \
  and\ \bibinfo {author} {\bibfnamefont {R.~v.}\ \bibnamefont {Leeuwen}},\
  }\href@noop {} {\bibfield  {journal} {\bibinfo  {journal} {J. Chem. Phys.}\
  }\textbf {\bibinfo {volume} {130}},\ \bibinfo {pages} {224101} (\bibinfo
  {year} {2009})}\BibitemShut {NoStop}%
\bibitem [{\citenamefont {Balzer}\ and\ \citenamefont
  {Bonitz}(2012)}]{balzer_nonequilibrium_2012}%
  \BibitemOpen
  \bibfield  {author} {\bibinfo {author} {\bibfnamefont {K.}~\bibnamefont
  {Balzer}}\ and\ \bibinfo {author} {\bibfnamefont {M.}~\bibnamefont
  {Bonitz}},\ }\href@noop {} {\emph {\bibinfo {title} {Nonequilibrium {Green}'s
  {Functions} {Approach} to {Inhomogeneous} {Systems}}}}\ (\bibinfo
  {publisher} {Springer},\ \bibinfo {year} {2012})\BibitemShut {NoStop}%
\bibitem [{\citenamefont {Stefanucci}\ and\ \citenamefont
  {Leeuwen}(2013)}]{stefanucci_nonequilibrium_2013}%
  \BibitemOpen
  \bibfield  {author} {\bibinfo {author} {\bibfnamefont {G.}~\bibnamefont
  {Stefanucci}}\ and\ \bibinfo {author} {\bibfnamefont {R.~v.}\ \bibnamefont
  {Leeuwen}},\ }\href@noop {} {\emph {\bibinfo {title} {Nonequilibrium
  {Many}-{Body} {Theory} of {Quantum} {Systems}: {A} {Modern}
  {Introduction}}}}\ (\bibinfo  {publisher} {Cambridge University Press},\
  \bibinfo {year} {2013})\BibitemShut {NoStop}%
\bibitem [{\citenamefont {Bernevig}\ \emph {et~al.}(2006)\citenamefont
  {Bernevig}, \citenamefont {Hughes},\ and\ \citenamefont
  {Zhang}}]{bernevig_quantum_2006}%
  \BibitemOpen
  \bibfield  {author} {\bibinfo {author} {\bibfnamefont {B.~A.}\ \bibnamefont
  {Bernevig}}, \bibinfo {author} {\bibfnamefont {T.~L.}\ \bibnamefont
  {Hughes}}, \ and\ \bibinfo {author} {\bibfnamefont {S.-C.}\ \bibnamefont
  {Zhang}},\ }\href@noop {} {\bibfield  {journal} {\bibinfo  {journal}
  {Science}\ }\textbf {\bibinfo {volume} {314}},\ \bibinfo {pages} {1757}
  (\bibinfo {year} {2006})}\BibitemShut {NoStop}%
\bibitem [{\citenamefont {{Hung}}\ \emph {et~al.}(2013)\citenamefont {{Hung}},
  \citenamefont {{Gurarie}},\ and\ \citenamefont
  {{Chin}}}]{QuenchWeakInteraction2013}%
  \BibitemOpen
  \bibfield  {author} {\bibinfo {author} {\bibfnamefont {C.-L.}\ \bibnamefont
  {{Hung}}}, \bibinfo {author} {\bibfnamefont {V.}~\bibnamefont {{Gurarie}}}, \
  and\ \bibinfo {author} {\bibfnamefont {C.}~\bibnamefont {{Chin}}},\
  }\href@noop {} {\bibfield  {journal} {\bibinfo  {journal} {Science}\ }\textbf
  {\bibinfo {volume} {341}},\ \bibinfo {pages} {1213} (\bibinfo {year}
  {2013})}\BibitemShut {NoStop}%
\bibitem [{\citenamefont {{Makotyn}}\ \emph {et~al.}(2014)\citenamefont
  {{Makotyn}}, \citenamefont {{Klauss}}, \citenamefont {{Goldberger}},
  \citenamefont {{Cornell}},\ and\ \citenamefont {{Jin}}}]{QuenchUnitary2014}%
  \BibitemOpen
  \bibfield  {author} {\bibinfo {author} {\bibfnamefont {P.}~\bibnamefont
  {{Makotyn}}}, \bibinfo {author} {\bibfnamefont {C.~E.}\ \bibnamefont
  {{Klauss}}}, \bibinfo {author} {\bibfnamefont {D.~L.}\ \bibnamefont
  {{Goldberger}}}, \bibinfo {author} {\bibfnamefont {E.~A.}\ \bibnamefont
  {{Cornell}}}, \ and\ \bibinfo {author} {\bibfnamefont {D.~S.}\ \bibnamefont
  {{Jin}}},\ }\href@noop {} {\bibfield  {journal} {\bibinfo  {journal} {Nature
  Physics}\ }\textbf {\bibinfo {volume} {10}},\ \bibinfo {pages} {116}
  (\bibinfo {year} {2014})}\BibitemShut {NoStop}%
\bibitem [{\citenamefont {Aoki}\ \emph {et~al.}(2014)\citenamefont {Aoki},
  \citenamefont {Tsuji}, \citenamefont {Eckstein}, \citenamefont {Kollar},
  \citenamefont {Oka},\ and\ \citenamefont
  {Werner}}]{aoki_nonequilibrium_2014}%
  \BibitemOpen
  \bibfield  {author} {\bibinfo {author} {\bibfnamefont {H.}~\bibnamefont
  {Aoki}}, \bibinfo {author} {\bibfnamefont {N.}~\bibnamefont {Tsuji}},
  \bibinfo {author} {\bibfnamefont {M.}~\bibnamefont {Eckstein}}, \bibinfo
  {author} {\bibfnamefont {M.}~\bibnamefont {Kollar}}, \bibinfo {author}
  {\bibfnamefont {T.}~\bibnamefont {Oka}}, \ and\ \bibinfo {author}
  {\bibfnamefont {P.}~\bibnamefont {Werner}},\ }\href@noop {} {\bibfield
  {journal} {\bibinfo  {journal} {Rev. Mod. Phys.}\ }\textbf {\bibinfo {volume}
  {86}},\ \bibinfo {pages} {779} (\bibinfo {year} {2014})}\BibitemShut
  {NoStop}%
\bibitem [{\citenamefont {Balzer}\ \emph {et~al.}(2016)\citenamefont {Balzer},
  \citenamefont {Schl\"{u}nzen},\ and\ \citenamefont
  {Bonitz}}]{balzer_stopping_2016}%
  \BibitemOpen
  \bibfield  {author} {\bibinfo {author} {\bibfnamefont {K.}~\bibnamefont
  {Balzer}}, \bibinfo {author} {\bibfnamefont {N.}~\bibnamefont
  {Schl\"{u}nzen}}, \ and\ \bibinfo {author} {\bibfnamefont {M.}~\bibnamefont
  {Bonitz}},\ }\href@noop {} {\bibfield  {journal} {\bibinfo  {journal} {Phys.
  Rev. B}\ }\textbf {\bibinfo {volume} {94}},\ \bibinfo {pages} {245118}
  (\bibinfo {year} {2016})}\BibitemShut {NoStop}%
\bibitem [{\citenamefont {Sch\"{u}ler}\ and\ \citenamefont
  {Pavlyukh}(2018)}]{schuler_spectral_2018}%
  \BibitemOpen
  \bibfield  {author} {\bibinfo {author} {\bibfnamefont {M.}~\bibnamefont
  {Sch\"{u}ler}}\ and\ \bibinfo {author} {\bibfnamefont {Y.}~\bibnamefont
  {Pavlyukh}},\ }\href@noop {} {\bibfield  {journal} {\bibinfo  {journal}
  {Phys. Rev. B}\ }\textbf {\bibinfo {volume} {97}},\ \bibinfo {pages} {115164}
  (\bibinfo {year} {2018})}\BibitemShut {NoStop}%
\bibitem [{\citenamefont {Lipavsky}\ \emph {et~al.}(1986)\citenamefont
  {Lipavsky}, \citenamefont {Spicka},\ and\ \citenamefont
  {Velicky}}]{lipavsky_generalized_1986}%
  \BibitemOpen
  \bibfield  {author} {\bibinfo {author} {\bibfnamefont {P.}~\bibnamefont
  {Lipavsky}}, \bibinfo {author} {\bibfnamefont {V.}~\bibnamefont {Spicka}}, \
  and\ \bibinfo {author} {\bibfnamefont {B.}~\bibnamefont {Velicky}},\
  }\href@noop {} {\bibfield  {journal} {\bibinfo  {journal} {Phys. Rev. B}\
  }\textbf {\bibinfo {volume} {34}},\ \bibinfo {pages} {6933} (\bibinfo {year}
  {1986})}\BibitemShut {NoStop}%
\bibitem [{Note1()}]{Note1}%
  \BibitemOpen
  \bibinfo {note} {The GKBA reduces the orginial KBEs, which constitute the a
  set of two-time integro-differential equations with memory, to a single-time
  non-Markovian equation.}\BibitemShut {Stop}%
\bibitem [{\citenamefont {Latini}\ \emph {et~al.}(2014)\citenamefont {Latini},
  \citenamefont {Perfetto}, \citenamefont {Uimonen}, \citenamefont {van
  Leeuwen},\ and\ \citenamefont {Stefanucci}}]{latini_charge_2014}%
  \BibitemOpen
  \bibfield  {author} {\bibinfo {author} {\bibfnamefont {S.}~\bibnamefont
  {Latini}}, \bibinfo {author} {\bibfnamefont {E.}~\bibnamefont {Perfetto}},
  \bibinfo {author} {\bibfnamefont {A.-M.}\ \bibnamefont {Uimonen}}, \bibinfo
  {author} {\bibfnamefont {R.}~\bibnamefont {van Leeuwen}}, \ and\ \bibinfo
  {author} {\bibfnamefont {G.}~\bibnamefont {Stefanucci}},\ }\href@noop {}
  {\bibfield  {journal} {\bibinfo  {journal} {Phys. Rev. B}\ }\textbf {\bibinfo
  {volume} {89}},\ \bibinfo {pages} {075306} (\bibinfo {year}
  {2014})}\BibitemShut {NoStop}%
\bibitem [{\citenamefont {Perfetto}\ \emph {et~al.}(2015)\citenamefont
  {Perfetto}, \citenamefont {Uimonen}, \citenamefont {van Leeuwen},\ and\
  \citenamefont {Stefanucci}}]{perfetto_first-principles_2015}%
  \BibitemOpen
  \bibfield  {author} {\bibinfo {author} {\bibfnamefont {E.}~\bibnamefont
  {Perfetto}}, \bibinfo {author} {\bibfnamefont {A.-M.}\ \bibnamefont
  {Uimonen}}, \bibinfo {author} {\bibfnamefont {R.}~\bibnamefont {van
  Leeuwen}}, \ and\ \bibinfo {author} {\bibfnamefont {G.}~\bibnamefont
  {Stefanucci}},\ }\href@noop {} {\bibfield  {journal} {\bibinfo  {journal}
  {Phys. Rev. A}\ }\textbf {\bibinfo {volume} {92}},\ \bibinfo {pages} {033419}
  (\bibinfo {year} {2015})}\BibitemShut {NoStop}%
\bibitem [{\citenamefont {Perfetto}\ \emph {et~al.}(2018)\citenamefont
  {Perfetto}, \citenamefont {Sangalli}, \citenamefont {Marini},\ and\
  \citenamefont {Stefanucci}}]{perfetto_ultrafast_2018}%
  \BibitemOpen
  \bibfield  {author} {\bibinfo {author} {\bibfnamefont {E.}~\bibnamefont
  {Perfetto}}, \bibinfo {author} {\bibfnamefont {D.}~\bibnamefont {Sangalli}},
  \bibinfo {author} {\bibfnamefont {A.}~\bibnamefont {Marini}}, \ and\ \bibinfo
  {author} {\bibfnamefont {G.}~\bibnamefont {Stefanucci}},\ }\href@noop {}
  {\bibfield  {journal} {\bibinfo  {journal} {J. Phys. Chem. Lett.}\ }\textbf
  {\bibinfo {volume} {9}},\ \bibinfo {pages} {1353} (\bibinfo {year}
  {2018})}\BibitemShut {NoStop}%
\bibitem [{\citenamefont {Bostr\"{o}m}\ \emph {et~al.}(2018)\citenamefont
  {Bostr\"{o}m}, \citenamefont {Mikkelsen}, \citenamefont {Verdozzi},
  \citenamefont {Perfetto},\ and\ \citenamefont
  {Stefanucci}}]{bostrom_charge_2018}%
  \BibitemOpen
  \bibfield  {author} {\bibinfo {author} {\bibfnamefont {E.~V.~n.}\
  \bibnamefont {Bostr\"{o}m}}, \bibinfo {author} {\bibfnamefont
  {A.}~\bibnamefont {Mikkelsen}}, \bibinfo {author} {\bibfnamefont
  {C.}~\bibnamefont {Verdozzi}}, \bibinfo {author} {\bibfnamefont
  {E.}~\bibnamefont {Perfetto}}, \ and\ \bibinfo {author} {\bibfnamefont
  {G.}~\bibnamefont {Stefanucci}},\ }\href@noop {} {\bibfield  {journal}
  {\bibinfo  {journal} {Nano Lett.}\ }\textbf {\bibinfo {volume} {18}},\
  \bibinfo {pages} {785} (\bibinfo {year} {2018})}\BibitemShut {NoStop}%
\bibitem [{\citenamefont {{Diehl}}\ \emph {et~al.}(2011)\citenamefont
  {{Diehl}}, \citenamefont {{Rico}}, \citenamefont {{Baranov}},\ and\
  \citenamefont {{Zoller}}}]{Diehl2011}%
  \BibitemOpen
  \bibfield  {author} {\bibinfo {author} {\bibfnamefont {S.}~\bibnamefont
  {{Diehl}}}, \bibinfo {author} {\bibfnamefont {E.}~\bibnamefont {{Rico}}},
  \bibinfo {author} {\bibfnamefont {M.~A.}\ \bibnamefont {{Baranov}}}, \ and\
  \bibinfo {author} {\bibfnamefont {P.}~\bibnamefont {{Zoller}}},\ }\href@noop
  {} {\bibfield  {journal} {\bibinfo  {journal} {Nature Physics}\ }\textbf
  {\bibinfo {volume} {7}},\ \bibinfo {pages} {971} (\bibinfo {year}
  {2011})}\BibitemShut {NoStop}%
\bibitem [{\citenamefont {Bardyn}\ \emph {et~al.}(2018)\citenamefont {Bardyn},
  \citenamefont {Wawer}, \citenamefont {Altland}, \citenamefont
  {Fleischhauer},\ and\ \citenamefont {Diehl}}]{BardynPRX2018}%
  \BibitemOpen
  \bibfield  {author} {\bibinfo {author} {\bibfnamefont {C.-E.}\ \bibnamefont
  {Bardyn}}, \bibinfo {author} {\bibfnamefont {L.}~\bibnamefont {Wawer}},
  \bibinfo {author} {\bibfnamefont {A.}~\bibnamefont {Altland}}, \bibinfo
  {author} {\bibfnamefont {M.}~\bibnamefont {Fleischhauer}}, \ and\ \bibinfo
  {author} {\bibfnamefont {S.}~\bibnamefont {Diehl}},\ }\href@noop {}
  {\bibfield  {journal} {\bibinfo  {journal} {Phys. Rev. X}\ }\textbf {\bibinfo
  {volume} {8}},\ \bibinfo {pages} {011035} (\bibinfo {year}
  {2018})}\BibitemShut {NoStop}%
\bibitem [{Note2()}]{Note2}%
  \BibitemOpen
  \bibinfo {note} {In our case, the gap closing of $\protect \mathbf
  {h}_\protect \mathrm {aux}(\protect \mathbf {k};t)$ occurs at the $\Gamma $
  point. In the noninteracting system in equilibrium, $\protect \mathbf
  {h}_\protect \mathrm {aux}(\protect \mathbf {k})=\protect \mathbf
  {h}(\protect \mathbf {k})$, hence the gap closing is equivalent to
  $M=-2$.}\BibitemShut {Stop}%
\bibitem [{Note3()}]{Note3}%
  \BibitemOpen
  \bibinfo {note} {The index $C_t$ is identical to the Chern number $C$ of
  $\protect \mathbf {h}_\protect \mathrm {aux}$ in equilibrium.}\BibitemShut
  {Stop}%
\bibitem [{supplement()}]{supplement}%
  \BibitemOpen
  \bibinfo {note} {Supplemental material, available online.}\BibitemShut
  {Stop}%
\bibitem [{Note4()}]{Note4}%
  \BibitemOpen
  \bibinfo {note} {In the thermodynamic limit, $C_t(t)$ is identical to
  $C_\protect \mathrm {inst}(t)$. The pronounced finite-size effects for
  $V=0.65$ since the very weak dephasing leads to the deviations in Fig.~\ref
  {fig:chern} (c)--(d).}\BibitemShut {Stop}%
\bibitem [{\citenamefont {Budich}\ \emph {et~al.}(2013)\citenamefont {Budich},
  \citenamefont {Trauzettel},\ and\ \citenamefont
  {Sangiovanni}}]{budich_fluctuation-driven_2013}%
  \BibitemOpen
  \bibfield  {author} {\bibinfo {author} {\bibfnamefont {J.~C.}\ \bibnamefont
  {Budich}}, \bibinfo {author} {\bibfnamefont {B.}~\bibnamefont {Trauzettel}},
  \ and\ \bibinfo {author} {\bibfnamefont {G.}~\bibnamefont {Sangiovanni}},\
  }\href@noop {} {\bibfield  {journal} {\bibinfo  {journal} {Phys. Rev. B}\
  }\textbf {\bibinfo {volume} {87}},\ \bibinfo {pages} {235104} (\bibinfo
  {year} {2013})}\BibitemShut {NoStop}%
\bibitem [{Note5()}]{Note5}%
  \BibitemOpen
  \bibinfo {note} {This manifestation of the topological invariant is
  unaffected by the interaction~\cite{frohlich_gauge_2013}.}\BibitemShut
  {Stop}%
\bibitem [{\citenamefont {Schl\"{u}nzen}\ \emph {et~al.}(2017)\citenamefont
  {Schl\"{u}nzen}, \citenamefont {Joost}, \citenamefont {Heidrich-Meisner},\
  and\ \citenamefont {Bonitz}}]{schlunzen_nonequilibrium_2017}%
  \BibitemOpen
  \bibfield  {author} {\bibinfo {author} {\bibfnamefont {N.}~\bibnamefont
  {Schl\"{u}nzen}}, \bibinfo {author} {\bibfnamefont {J.-P.}\ \bibnamefont
  {Joost}}, \bibinfo {author} {\bibfnamefont {F.}~\bibnamefont
  {Heidrich-Meisner}}, \ and\ \bibinfo {author} {\bibfnamefont
  {M.}~\bibnamefont {Bonitz}},\ }\href@noop {} {\bibfield  {journal} {\bibinfo
  {journal} {Phys. Rev. B}\ }\textbf {\bibinfo {volume} {95}},\ \bibinfo
  {pages} {165139} (\bibinfo {year} {2017})}\BibitemShut {NoStop}%
\bibitem [{Note6()}]{Note6}%
  \BibitemOpen
  \bibinfo {note} {We choose $f(\tau ) = 10 \tau ^3-15\tau ^4+6\tau ^5$ for
  $\tau \in [0,1]$.}\BibitemShut {Stop}%
\bibitem [{\citenamefont {Sch\"{u}ler}\ \emph {et~al.}(2018)\citenamefont
  {Sch\"{u}ler}, \citenamefont {Murakami},\ and\ \citenamefont
  {Werner}}]{schuler_nonthermal_2018}%
  \BibitemOpen
  \bibfield  {author} {\bibinfo {author} {\bibfnamefont {M.}~\bibnamefont
  {Sch\"{u}ler}}, \bibinfo {author} {\bibfnamefont {Y.}~\bibnamefont
  {Murakami}}, \ and\ \bibinfo {author} {\bibfnamefont {P.}~\bibnamefont
  {Werner}},\ }\href@noop {} {\bibfield  {journal} {\bibinfo  {journal} {Phys.
  Rev. B}\ }\textbf {\bibinfo {volume} {97}},\ \bibinfo {pages} {155136}
  (\bibinfo {year} {2018})}\BibitemShut {NoStop}%
\bibitem [{\citenamefont {Alvermann}\ and\ \citenamefont
  {Fehske}(2011)}]{alvermann_high-order_2011}%
  \BibitemOpen
  \bibfield  {author} {\bibinfo {author} {\bibfnamefont {A.}~\bibnamefont
  {Alvermann}}\ and\ \bibinfo {author} {\bibfnamefont {H.}~\bibnamefont
  {Fehske}},\ }\href@noop {} {\bibfield  {journal} {\bibinfo  {journal} {J.
  Comput. Phys.}\ }\textbf {\bibinfo {volume} {230}},\ \bibinfo {pages} {5930}
  (\bibinfo {year} {2011})}\BibitemShut {NoStop}%
\bibitem [{\citenamefont {Schl\"{u}nzen}\ and\ \citenamefont
  {Bonitz}(2016)}]{schlunzen_nonequilibrium_2016}%
  \BibitemOpen
  \bibfield  {author} {\bibinfo {author} {\bibfnamefont {N.}~\bibnamefont
  {Schl\"{u}nzen}}\ and\ \bibinfo {author} {\bibfnamefont {M.}~\bibnamefont
  {Bonitz}},\ }\href@noop {} {\bibfield  {journal} {\bibinfo  {journal}
  {Contributions to Plasma Physics}\ }\textbf {\bibinfo {volume} {56}},\
  \bibinfo {pages} {5} (\bibinfo {year} {2016})}\BibitemShut {NoStop}%
\bibitem [{\citenamefont {Fr\"{o}hlich}\ and\ \citenamefont
  {Werner}(2013)}]{frohlich_gauge_2013}%
  \BibitemOpen
  \bibfield  {author} {\bibinfo {author} {\bibfnamefont {J.}~\bibnamefont
  {Fr\"{o}hlich}}\ and\ \bibinfo {author} {\bibfnamefont {P.}~\bibnamefont
  {Werner}},\ }\href@noop {} {\bibfield  {journal} {\bibinfo  {journal} {EPL
  (Europhysics Letters)}\ }\textbf {\bibinfo {volume} {101}},\ \bibinfo {pages}
  {47007} (\bibinfo {year} {2013})}\BibitemShut {NoStop}%
\end{thebibliography}

%

\end{document}